\documentclass[12pt]{article}\usepackage{cc_rg}

\begin{document}

\title{ An Application of \\
Renormalization Group Techniques\\
to \\
Classical Information Theory}

\author{Robert R. Tucci\\
        P.O. Box 226\\
        Bedford,  MA   01730\\
    tucci@ar-tiste.com}

\date{ \today}

\maketitle

\vskip2cm
\section*{Abstract}
We apply Renormalization Group (RG) techniques to
Classical Information Theory, in the limit of large codeword size $n$.
In particular, we apply RG techniques to (1) noiseless coding (i.e., a coding
used for compression) and (2) noisy coding (i.e., a coding used
for channel transmission). Shannon's ``first" and ``second" theorems
refer to (1) and (2), respectively.
Our RG technique
uses composition class (CC) ideas,
so we call our technique Composition Class Renormalization Group
(CCRG). Often, CC's  are called ``types" instead of CC's,
and their theory is referred to as the  ``Method of Types".
For (1) and (2), we find that the probability of error
can be expressed as an Error Function whose argument contains
variables that obey renormalization group equations. We describe a
computer program called WimpyRG-C1.0 that implements the ideas of
this paper. C++ source code for WimpyRG-C1.0 is publicly available.

\newpage
\section{Introduction}

Renormalization Group (RG) techniques \cite{Gold} are a panoply of
techniques that serve to obtain asymptotic limits. RG
techniques usually apply to a system with a very large number of
degrees of freedom that is described by a partition function $Z$.
Most RG techniques comprise an iterative step
(i.e., a step which is performed repeatedly)
consisting of a decimation followed by a rescaling.
Decimation involves reducing the number of degrees of
freedom. Rescaling involves rescaling the variables of $Z$ so as to bring
$Z$ to the same form  it had before the previous decimation.
(Curiously, in Roman times, the word ``decimate"
 meant to kill 1 out of every 10 prisoners. The modern
meaning of the word is more like killing 9 out of every 10).

In this paper, we apply RG techniques to Classical Information
Theory\cite{cover-thomas}\cite{blahut}
in the limit of large codeword size $n$.
In particular, we apply RG techniques to (1) noiseless coding (i.e., a coding
used for compression) and
(2) noisy coding (i.e., a coding used
for channel transmission).
Shannon's ``first" and ``second" theorems
refer to (1) and (2), respectively.
For (1), we consider the special case of
Csisz\'ar-K\"orner (CK) universal code.
For (2), we consider the special case of
random encoding and maximum-likelihood (ML) decoding.
For these special cases of (1) and (2),
we find that the probability of error
can be expressed as an Error Function
(see Appendix \ref{app:err-fun}) whose
argument contains variables
that obey RG equations.

Of course, there is no unique way of
applying RG techniques to Classical Information Theory.
The way shown in this paper is new, to our knowledge.
Our RG technique uses composition class (CC) ideas,
so we call our technique Composition Class Renormalization Group
(CCRG). Often, CC's  are called ``types" instead of CC's,
and their theory is referred to as the  ``Method of Types".

We end this paper by describing the internal algorithms
and typical input and output of a computer program
called WimpyRG-C1.0 that implements
the ideas of this paper. (The 1.0 is the version number.
The C before the 1.0 stands
for ``Classical", to distinguish this program from
a Q (Quantum) version of WimpyRG that we expect to deliver in the future.)
C++ source code for WimpyRG-C1.0 is publicly available,
at www.ar-tiste.com/WimpyRG.html .

This paper straddles two fields (RG and Classical Information
Theory) which are seldom used together within previous literature.
It is therefore most likely that the reader is not closely
acquainted with both of these fields. To help readers acquainted
with only one of these two fields, the author has strived to make
this paper as self-contained as reasonably possible.

Before embarking on long, complicated calculations, let us discuss
a simple example that illustrates the manner in which we will
apply RG ideas to Information Theory
in this paper.

We show in this paper that
the probability of error for both noiseless
and noisy coding can be expressed
as an integral of the following type:

\beq
I = \int_{\xi}^{+\infty}dx \; e^{-n f(x)}
\;,
\eeq
where $n>>1$.
Suppose
$f: Reals\rarrow Reals$ is a convex (i.e., shaped like a cup $\cup$) function with a
minimum at $x_0$. Let $\Delta x = x - x_0$, $\Delta \xi = \xi -
x_0$, and $ F(\Delta x) = f(x) $. Then $I$ can be rewritten as

\beq
I = \int_{\Delta \xi}^{+\infty}d\Delta x \;
e^{-n F(\Delta x)}
\;.
\label{eq:$I$-def2}
\eeq

$I$ can be approximated as follows

\beq
I
\approx e^{-n F(\Delta \xi)}
\;.
\label{eq:laplace-leading}
\eeq
This approximation for $I$ is
the leading term of an asymptotic expansion.
This method of obtaining asymptotic expansions of integrals is
usually called
{\it Laplace's Method} \cite{asymp}, named after the inventor of
the closely related Laplace Transform.
Unfortunately, the $I$-approximation given by Eq.(\ref{eq:laplace-leading})
is poor for those $\Delta \xi_0$ for which
$F(\Delta \xi_0)=0$. Indeed, $e^{-n F(\Delta \xi_0)}$ is indeterminate because
$n F(\Delta \xi_0) = \infty \cdot 0$.
Our goal is to devise an $I$-approximation  that
overcomes this limitation.

Suppose, for example, that $F$
is quadratic in $\Delta x$:

\beq
F(\Delta x) = \frac{a}{2} (\Delta x)^2
\;,
\label{eq:quad-F}
\eeq
for some $a>0$. Then we can do the integration
in Eq.(\ref{eq:$I$-def2}) exactly
in terms of Error Functions (see Appendix \ref{app:err-fun})

\beqa
I &=& \int^{+\infty}_{\Delta \xi}
d\Delta x \;  e^{-n \frac{a}{2} (\Delta x)^2}\\
&=& \sqrt{\frac{\pi}{2na}}
\erfc \left( \Delta \xi \sqrt{ \frac{na}{2}} \right)
\;.
\eeqa
Using RG ideas, we can
generalize this result, valid only for
a quadratic $F$, to more general types of $F$.
In Eq.(\ref{eq:$I$-def2}),
let us rescale the parameters
$\Delta \xi, n$ and the integration variable
$\Delta x$, but keep the value of
$I$ fixed. Then
\beq
I = \int_{\Delta \xi^\wedge}^{+\infty}d\Delta x^\wedge\; J
\;e^{-n^\wedge F^\wedge(\Delta x)}
\;,
\eeq
where $J$ is a Jacobian, and where, for some parameter $s>0$, we define

\beq
n^\wedge = e^{s} n
\;,
\eeq
and

\beq
F^\wedge(\Delta x) =F(\Delta x^\wedge) = e^{-s}F(\Delta x)
\;.
\eeq

For $s=\delta s$ where $0<\delta s<<1$, we get:

\beq
\delta s = \frac{-\delta F}{F}
\;.\label{eq:dF-ds}
\eeq
From Eq.(\ref{eq:dF-ds}), we get the following
``RG Equation":

\beq
\frac{d\Delta \xi^\sfun}{ds} = \frac{-F(\Delta \xi^\sfun)}{F_1(\Delta \xi^\sfun)}
\;,\label{eq:x$I$-rg-eq}
\eeq
where $F_n$ is the $n$th derivative of $F$,
and we have replaced the symbol $\wedge$ by $(s)$.
Of course, this RG equation is trivial and can be solved immediately:

\beq
\Delta \xi^\sfun = F^{-1}(e^{-s}F(\Delta \xi))
\;.\label{eq:x$I$-rg-eq-sol}
\eeq
In the more complicated examples presented later in this paper,
one gets a system of coupled RG equations with complicated
boundary conditions. Such systems of RG equations
usually cannot be solved exactly, but they can be solved
numerically with a computer.

We can calculate the Jacobian $J$ as follows:

\beq
\Delta x^{(\delta s)} = \Delta x + \delta \Delta x = \Delta x
-\delta s\; \frac{F}{F_1}
\;,
\eeq
so

\beq
J^{-1} = \left |
\pder{\Delta x^{(\delta s)}}{\Delta x} \right|
= \left |
1 - \delta s\left( 1 - \frac{F F_2}{(F_1)^2}\right)
\right|
\label{eq:jacobian}
\;.
\eeq
Note that  we are justified in setting
$J \approx 1$ if we
are only interested in finding $I$ to leading order in $n$.

Suppose $\Delta
\xi>0$. Since $F(\Delta \xi)$ is a convex function with minimum at
the origin, as $s$  increases (and therefore also $n$ increases),
then, according to Eq.(\ref{eq:x$I$-rg-eq}), $\Delta \xi$ decreases.
Likewise, if $\Delta \xi<0$, then as $s$ increases, $\Delta \xi$
increases. In both cases, $\Delta \xi$ is attracted to zero as $s$ increases.
By making $s$ large enough, we can make $\Delta\xi$ small enough
so that $F$ is well approximated by its quadratic approximation:

\begin{subequations}\label{eq:rg-approx}
\begin{eqnarray}
I &=&
\int_{\Delta \xi^\sfun}^{+\infty} d\Delta x^\sfun \;
J\; e^{-n^\sfun F^\sfun (\Delta x)}\\
&\approx&
e^{-n^\sfun  F(0)}
\int_{\Delta \xi^\sfun}^{+\infty} d\Delta x^\sfun \;
 e^{-n^\sfun \frac{F_2(0)}{2} (\Delta x)^2}\\
&\approx&
e^{-n^\sfun  F(0)}
 \sqrt{ \frac{\pi}{2 n^\sfun F_2(0)}}
 \erfc\left( \Delta \xi^\sfun \sqrt{ \frac{n^\sfun F_2(0)}{2}} \right)
\label{eq:rg-approx-final}
\;.
\eeqa
In Eq.(\ref{eq:rg-approx}), to go from
line (a) to (b), we replaced $F$ by its Taylor expansion up
to second order (this is valid for very large $s$) and we approximated $J$ by one
(this is valid to leading order in $n$).
Eq.(\ref{eq:rg-approx-final}) is typical of the type of
approximations that we propose in this paper.

Before leaving our toy example, it is instructive
to compare the $I$-approximation Eq.(\ref{eq:rg-approx-final})
to the exact answer in case $F$ is
quadratic. So assume $F(0) = 0$
and $F_2(0)=a$ as in Eq.(\ref{eq:quad-F}).
For such an $F$, one can show
from Eq.(\ref{eq:x$I$-rg-eq-sol}) that

\beq
\Delta \xi^\sfun = e^{\frac{-s}{2}} \Delta \xi
\;.
\eeq
Furthermore, one can show from Eq.(\ref{eq:jacobian}) that

\beq
J^{-1} = e^{\frac{-s}{2}}
\;.
\eeq
By definition,

\beq
n^\sfun = e^{s} n
\;.
\eeq
Thus,

\beqa
I &=&
\int_{\Delta \xi^\sfun}^{+\infty} d\Delta x^\sfun \;
J \;e^{-n^\sfun F^\sfun (\Delta x)}\\
&=&
e^{\frac{s}{2}} \sqrt{ \frac{\pi}{2 n^\sfun a}}
 \erfc\left( \Delta \xi^\sfun \sqrt{ \frac{n^\sfun a}{2}} \right)\\
&=&
\sqrt{ \frac{\pi}{2 n a}}
 \erfc\left( \Delta \xi \sqrt{ \frac{n a}{2}} \right)
\;.
\eeqa
Hence, we see that for a quadratic $F$,
$I$-approximation Eq.(\ref{eq:rg-approx-final})
differs from the exact answer by a factor of
$e^{\frac{s}{2}}$. This discrepancy is due to
the fact that we neglected the Jacobian in deriving
$I$-approximation Eq.(\ref{eq:rg-approx-final}).

\newpage
\section{Notation}
In this section, we will present some notation
that will be used throughout the paper.

RHS and LHS will stand for ``right hand side" and ``left hand side", respectively.
When we say ``$x$ (ditto, $y$) is $A$ (ditto, $B$)" we will
mean that $x$ is $A$ and $y$ is $B$.

The number of elements in a set $S$ will be denoted by $|S|$. Let
$Z_{a, b} = \{ a, a+1, a+2, \ldots , b\}$ for any integers $a\leq b$.
Let $x^{\#n}$ represent an n-tuple consisting of $n$ copies of
$x$. For example, $x^{\#3} = (x,x,x)$. Any of the following
notations will be used to denote a set with indexed elements $A_i$
where $i\in S_\rvi$: $\{A_i\}_{\forall i} =  \{ A_i : \forall i \}
= \{ A_i : \forall i \in S_\rvi\}$. Any of the following notations
will be used to denote an ordered set (or vector) with components
$A_i$: $\vec{A} = ( A_i )_{\forall i} =( A_i : \forall i)$. For
example, we might refer to a matrix with elements $A_{i,j}$ by
$[A_{i, j}]_{\forall (i, j)}$. The components of a  vector
$\vec{A}$ will be denoted by $\vec{A}_i =( A_i : \forall i)_i$.
For any function $f: S_\rvx\rarrow Reals$, let $\prodset{f(x)}{x}
= \prod_{x \in S_\rvx} f(x)$. We will sometimes abbreviate
$\prodset{f(x)}{x}$ by $\prod \{f\}$. If $\vec{x}\in S_\rvx^n$
represents an n-letter codeword, we reserve the upper index
location for the label of a letter in the codeword. Thus, we will
denote the codeword $\vec{x}$ also by $ x^{/n}$, and its $i$'th
component by $(\vec{x})^i = x^{i/n}\in S_\rvx$ for all $i\in Z_{1,
n}$. $Reals^{n\times m}$ will represent $n$ by $m$ matrices with
real entries.

Given two sequences of real numbers $(a_n)_{\forall n}$ and
$(b_n)_{\forall n}$ where $n\in Z_{1, \infty}$,
we will often write $a_n \approx b_n$
to mean that $\lim_{n\rarrow \infty} \frac{a_n}{b_n} = 1$.

$pd(S)$ will represent all probability distributions on $S$; that
is, all functions $P:S\rarrow [0, 1]$ such that $\sum_{x \in S}P(x) =1$.
Random variables will be denoted by underlining. The set of all
possible values that a random variable $\rvx$ can assume will be
denoted by $ S_\rvx$. Let $|S_\rvx| = N_\rvx$. For any $x\in
S_\rvx$, the probability $P(\rvx = x) = P_\rvx(x)$ often will be
abbreviated by $ P(x)$ if this will not lead to confusion.
Likewise, for two random variables $\rvx, \rvy$, $S_{\rvx, \rvy} =
S_\rvx \times S_\rvy = \{(x, y): x\in S_\rvx, y\in S_\rvy\}$ and
$N_{\rvx, \rvy} = |S_{\rvx, \rvy}|=N_\rvx N_\rvy$. The probability
$P(\rvx = x, \rvy =y ) = P_{\rvx, \rvy}(x, y)$ often will be
abbreviated by $P(x, y)$ if this will not lead to confusion.

For any statement $S$, let $\theta(S)$ denote the ``truth
function" or ``indicator function": it equals 1 if $S$ is true and
it equals 0 if  $S$ is false. For example, $\theta(x>0)$ is the
unit step function. The Kronecker delta function is defined as
$\delta(x, y) = \delta^y_x  = \theta(x=y)$. Its continuum version,
the Dirac delta function, is defined by

\beq
\delta(x) = \int_{-\infty}^{+\infty}\frac{dk}{2\pi} e^{i kx -\epsilon k^2}
\;,
\label{eq:dirac-del} \eeq
for some infinitesimal $\epsilon>0$.
The Dirac function $\delta(x)$
 has unit area:
$\int_{-\infty}^{+\infty} \delta(x) = 1$, and is
sharply peaked at $x=0$. The identity
$\delta(x) = \frac{d}{dx} \theta(x>0)$
is easily proven using the sharply peaked and unit area properties
of $\delta(x)$. This identity connecting the Dirac delta function
and the unit step function
leads us to suspect that there is an integral representation
for the unit step function, analogous to Eq.(\ref{eq:dirac-del})
for the Dirac delta function.
Indeed, there is. Suppose $K>0$. The following equation is easy to prove
using contour integration in the complex plane:
\beq
\theta(x>0) = \frac{1}{2 \pi i} \int_{K - i \infty}^{K + i\infty} \frac{dk}{k} e^{kx}
\;.\label{eq:theta-int-rep}
\eeq
See  Fig.\ref{fig:theta-complex-int}. For $x>0$,
the integration contour can be deformed
so that it wraps around the point $k=0$.
By integrating around this pole, it is easy to show that
for $x>0$, the
RHS of Eq.(\ref{eq:theta-int-rep}) equals 1.
For $x<0$, the integration contour can be deformed so that
it wraps around the point $k = + \infty$. Thus, for $x<0$, the RHS of
Eq.(\ref{eq:theta-int-rep}) equals 0.

\begin{figure}[h]
    \begin{center}
    \epsfig{file=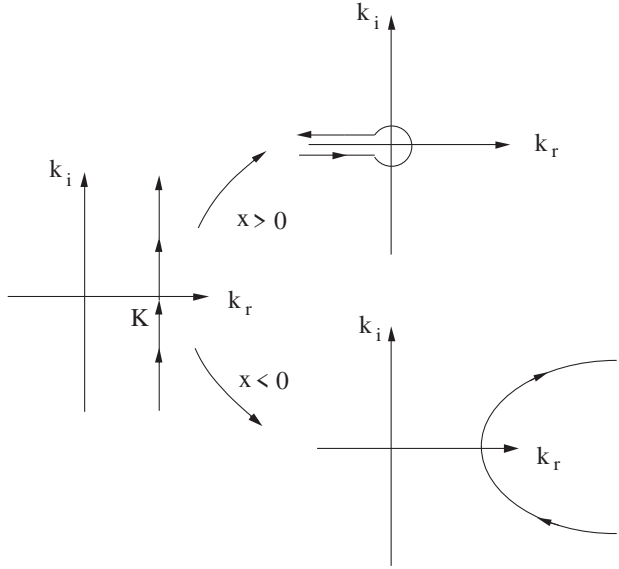, height=3.0in}
    \caption{For complex integral
    Eq.(\ref{eq:theta-int-rep}), one can deform the
contour of integration differently for $x<0$ and $x>0$.}
    \label{fig:theta-complex-int}
    \end{center}
\end{figure}

The {\it Shannon entropy} associated with the random variable $\rvx$
will be represented by any of the following:

\beq H_{P_\rvx}(\rvx)= H(P_\rvx) = H(\vec{P})= H(P(x))_{\forall x}
= - \sum_x P(x) \ln P(x)
\;.
\eeq
 Likewise, the {\it relative
entropy (also called the Kullback Liebler distance)} between two
probability distributions $P(x)$ and $Q(x)$ will be represented by
any of the following:

\beq D(P_\rvx//Q_\rvx) = D(\vec{P}// \vec{Q})=
D(P(x)//Q(x))_{\forall x} = \sum_x P(x) \ln \frac{P(x)}{Q(x)}
\;.
\eeq
We will also use the  {\it conditional entropy}

\beq
H(\rvx|\rvy) = -\sum_{x, y} P(x, y) \ln P(x|y)
\;,
\eeq
and the {\it mutual entropy}:

\beq
H(\rvx : \rvy) = \sum_{x, y} P(x, y) \ln \frac{P(x, y)}{P(x) P(y)}
\;.
\eeq
Note that we have defined our entropies in terms
 of base  $e$ rather than base 2 logs. Of course,
$\log_a (x) = \frac{\log_b x}{ \log_b a}$ so $\log_2 x= \frac{\ln x}{\ln 2}$

Let
$\calD P = \prodset{dP(x)}{x}$. For any function
$f: Reals^{N_\rvx} \rarrow Reals$, define

\beq
\int \calD P\; f(P) = \prodset{\int_{-\infty}^\infty dP(x)}{x} f(P)
\;,
\eeq
and

\beq
\int_{pd(S_\rvx)} \calD P\; f(P) = \int\calD P \; \theta(P\geq 0) \delta( \sum_x P(x) -1) f(P)
\;.
\eeq
It is easy to prove by induction that

\beq
\int_{pd(S_\rvx)} \calD P\; 1
 = \frac{1}{(N_\rvx -1)!}
\;.
\eeq

\section{Composition Classes}
In this section we will discuss composition classes (CC's). Often,
CC's  are called ``types" instead of CC's,
and their theory is
referred to as the  ``Method of Types". The term ``type" is very vague,
so  we will shun it, and use the more
specific term CC. This section reviews and extends standard material
on CC's as found in, for example, the books by Cover and Thomas
\cite{cover-thomas} and the one by
 Blahut \cite{blahut}.

In the mathematical theory of Statistics, one
often considers a sequence of $n$  random
variables $(\rvx^1, \rvx^2, \ldots \rvx^n) = \rvx^{/n}= \vec{\rvx}\in S_\rvx^n$.
Information Theory  also deals with such sequences, where they are
called a {\it word} (or {\it codeword} or {\it block}) of {\it letters}
(or {\it symbols}) $x$ from the {\it alphabet} $S_\rvx$.
We will assume the simplest case, wherein the $n$ random variables are
independent, identically distributed ({\it i.i.d.}), and
each $\rvx^i$ is distributed (``drawn") according to a probability distribution $Q:S_\rvx \rarrow [0,1]$.
In what follows, we will often refer to $Q$ as the {\it Center of Mass (CM) probability distribution},
(The reason for this name will be explained later.)

Let $n(x|\vec{x})$ represent the number of times that the letter $x$ occurs in the
word $\vec{x}$.
A {\it composition class} $\class{x}$ (also called  a ``type"
or ``empirical distribution" or ``relative frequency")
is defined by

\beq
\class{x} =\{ \vec{y} : \forall x \in S_\rvx, \;n(x|\vec{x}) = n(x|\vec{y})\}
\;.
\eeq
Clearly, this defines an equivalence relation on (and a disjoint partition for)
 the set $S^n_\rvx$. To each CC, there corresponds
a probability distribution given by

\beq
P_{\class{x}}(x) = \frac{n(x|\vec{x})}{n}
\;
\eeq
for all $x\in S_\rvx$.
In the notation $\class{x}$, the
CC is  specified by giving one of its elements $\vec{x}$.
Alternatively, one can specify a CC by  giving its probability distribution:

\beq
\cc(P) = \{ \vec{x} : \pclass{x} = P\}
\;.
\eeq
Hence
$\cc(\pclass{y}) = \class{y}$.

Define
$\tworow {S^n_\rva}{S^n_\rvb}$
to be the set of all $2\times n$ matrices
$\tworow {\vec{a}}{\vec{b}}$, where $\vec{a}\in S_\rva^n$ and
$\vec{b}\in S_\rvb^n$ are $n$-dimensional row vectors.
For some  $x^{/n}\in \tworow {S^n_\rva}{S^n_\rvb}$,
the  CC
denoted by $\cc(x^{/n})= \classtwo{a}{b}$ is defined as before, as
the set of all $2\times n$ matrices  $y^{/n}\in \tworow {S^n_\rva}{S^n_\rvb}$
 such that, for all
column vectors $x = \tworow{a}{b}$ with $a\in S_\rva$ and $b\in
S_\rvb$, one has $n(x|y^{/n} ) = n(x|x^{/n} )$.

For any $A\subset S_\rvx^n$,
it is convenient to define the following two sets:

\beq
\calC(A) = \{ \class{x} : \forall \vec{x} \in A\}
\;,
\eeq

\beq
\calP(A) = \{\pclass{x} : \forall \vec{x} \in A \}\subset pd(S_\rvx)
\;.
\eeq
Note that these two sets are in 1-1 correspondence. For $A= S^n_\rvx$,
they become
$\calC(S_\rvx^n)$ and  $\calP(S_\rvx^n)$.

For large $n$, we can easily estimate the number
of elements in a CC and the number of
$C(\vec{x})$ for all $\vec{x}\in S^n_\rvx$.

\begin{claim} As $n\rarrow \infty$,
\beq
|\class{x}| \approx
\frac{\exp\left[n H(\pclass{x}) \right]}{
(2\pi n)^{\frac{1}{2} (N_\rvx -1)}
\sqrt{ \prod\{ \pclass{x}\}}
}
\;,
\label{eq:approx-card-cc}
\eeq
and

\beq
|\calC(S_\rvx^n)|
\approx
\frac{n^{N_\rvx -1 }}{(N_\rvx -1 )!}
\;.
\label{eq:approx-num-ccs}
\eeq
\end{claim}
{\bf proof:}

The exact number of elements in $\class{x}$ is given by

\beq |\class{x}| = \frac{n!}{ \prod\{ n(x|\vec{x})!\}_{\forall x}
} \;. \label{eq:exact-card-cc} \eeq Recall the first term of
Stirling's asymptotic expansion, for large
$n$, of the factorial $n!$ :

\beq
n! \approx \sqrt{2 \pi n} \;\;e^{-n} n^n
\;.
\eeq
Applying this approximation to the factorials in
Eq.(\ref{eq:exact-card-cc}) immediately yields
Eq.(\ref{eq:approx-card-cc}).
Ref.\cite{cover-thomas} proves
that
$| \class{x} |$
is bounded below and above as follows:

\beq
\frac{1}{(n+1)^{N_\rvx}} \leq | \class{x} |
 \leq \exp\left[n H(\pclass{x}) \right]
\;.
\eeq

Since
$\pclass{x} = \frac{1}{n} (n_1, n_2,\ldots , n_{N_\rvx})$ ,
where $n_1, n_2,\ldots , n_{N_\rvx}\in Z_{0, n}$,
it follows that the exact number of CC's in
$pd(S_\rvx)$ is given by

\beq | \calC(S_\rvx^n)| = \sum_{n_1=0}^n \sum_{n_2=0}^n \cdots
\sum_{n_{N_\rvx}=0}^n \delta( \sum_{j=1}^{N_\rvx} n_j , n) \;.
\label{eq:exact-num-ccs} \eeq The previous equation immediately
implies that

\beq
 |\calC(S_\rvx^n)| \leq (n+1)^{N_\rvx}
\;.
\eeq
Suppose $f: Reals \rarrow Reals$.
For large $n$:

\beq
\sum_{k=0}^n f(k) \approx \int_0^{n+1} dk \; f(k)
\;.
\eeq
For any $n$:

\beq
\sum_{k=0}^n \delta(k, k_0) = \theta(0 \leq k_0 \leq n) = \int_0^n dk\; \delta(k-k_0)
\;.
\eeq
We can use the previous two equations
 to approximate all sums in Eq.(\ref{eq:exact-num-ccs})
by integrals. This  yields:

\beqa
|\calC(S_\rvx^n)|
&\approx &
\int_0^{n+1}dn_1\;
\int_0^{n+1}dn_2\;
\cdots
\int_0^{n+1}dn_{N_\rvx}\;
\delta(\sum_{j=1}^{N_\rvx} n_j - n)\\
&\approx&
n^{N_\rvx -1 }
\int_0^{1}dP_1\;
\int_0^{1}dP_2\;
\cdots
\int_0^{1}dP_{N_\rvx}\;
\delta(\sum_{j=1}^{N_\rvx} P_j - 1)\\
&\approx&
\frac{n^{N_\rvx -1 }}{(N_\rvx -1 )!}
\;.
\eeqa
QED

Let $Q(\vec{x})$ stand for the joint probability of the components of $\vec{\rvx}$.
Since we will assume that these components are i.i.d.,

\beq
Q(\vec{x}) = \prod\{ Q(x^i)\}_{\forall i}
\;.
\eeq
For $A\subset S^n_\rvx$, let

\beq
Q(A) = \sum_{\vec{x}\in A} Q(\vec{x})
\;.
\eeq
$Q(\vec{x}) $ can be expressed in terms of relative entropy as follows:

\beqa
Q(\vec{x})
&=& \prod\{ Q(x)^{n(x| \vec{x})} \}_{\forall x}\\
&=& \exp\left[n \sum_x \pclass{x}(x) \ln Q(x)\right]\\
&=& \exp\left[ -n H(\pclass{x}) - nD(\pclass{x}//Q)\right]
\;.
\eeqa
Combining this expression for $Q(\vec{x})$
with the approximation Eq.(\ref{eq:approx-card-cc})
for $|\class{x}|$ yields

\begin{subequations}\label{eq:q-c-vecx}
\begin{eqnarray}
Q(\class{x}) &= & |\class{x}| Q(\vec{x}) \\
&\approx&
\frac{
\exp\left[-n D(\pclass{x}//Q) \right]
}{
(2\pi n)^{\frac{1}{2} (N_\rvx -1)}
\sqrt{ \prod\{ \pclass{x}\}}
}
\;.
\eeqa

\begin{figure}[h]
    \begin{center}
    \epsfig{file=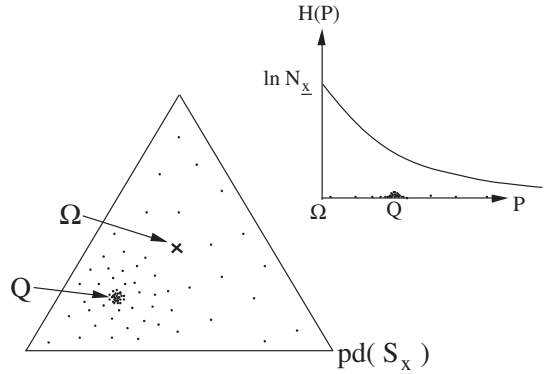, height=2.0in}
    \caption{Probability simplex $pd(S_\rvx)$ for $N_\rvx=3$.
    Two especially important points of the simplex
    are its geometric center $\Omega$ and its center of mass (CM) $Q$.
    The graph on the right illustrates how the entropy $H(P)$ decreases
    monotonically as point $P$ moves from the geometric center to the edges.}
    \label{fig:pd-Sx}
    \end{center}
\end{figure}

The set $pd(S_\rvx)$
is in 1-1 correspondence with a simplex in space $Reals^{N_\rvx}$.
For example, for $N_\rvx=3$, this probability simplex is the
region of $Reals^3$ that connects the corners $(1, 0, 0)$, $(0, 1, 0)$ and
$(0, 0, 1)$. Fig.\ref{fig:pd-Sx}
shows $pd(S_\rvx)$ for $N_\rvx=3$.
The probability distributions $\pclass{x}$ form a
finite subset of this simplex. In Fig.\ref{fig:pd-Sx},
the $\pclass{x}$ are represented by
dots inside $pd(S_\rvx)$ .
Other notable points of $pd(S_\rvx)$
are its
{\it geometric center} $\Omega = \left(\frac{1}{N_\rvx}\right)^{\#n}$
and the CM distribution
$Q(x)$.
From Eq.(\ref{eq:approx-card-cc})
it follows that the closer a CC is to the
geometric center $\Omega$, the more elements the CC has.
If we represent CC's by points of $pd(S^n_\rvx)$
with varying diameters, where fatter points represent
CC's with more elements, then the diameter of the points decreases
as we travel away from $\Omega$.
From Eq.(\ref{eq:q-c-vecx}),
it follows that the closer a
CC is to the CM distribution $Q(x)$,
the more probable the CC is.
As in Fig.\ref{fig:pd-Sx}, if we
show only the most probable CC's, then most of the CC's shown
cluster around the point $Q(x)$
(This is why we call $Q(x)$ and
$Q(\vec{x})$ the CM distribution.)

As mentioned in the introduction,
 most RG methods comprise an iterative step,
 (i.e., a step that is performed repeatedly)
consisting of
a {\bf decimation} followed by a {\bf rescaling}.
CCRG is slightly different from this.
In CCRG,
we perform a {\bf preliminary reduction} that
reduces a very large (i.e., infinite as $n\rarrow \infty$) number of degrees
of freedom to a small, fixed (i.e.,  $n$ independent)
number.
This is accomplished by
replacing sums like $\sum_{\vec{x}}$,
that run over $n$ discrete degrees of freedom,
by integrals like $\int \calD P_\rvx$,
that run over the far fewer $N_\rvx$ continuous
degrees of freedom that specify a point of $pd(S_\rvx)$.
After this preliminary reduction,
we perform an {\bf iterative step} consisting of
an infinitesimal  rescaling of $n$
 followed by a rescaling of
all other parameters in such a way that the form of the partition function
is not changed by the  iterative step.

The following two claims embody the
preliminary reduction step of CCRG.

\begin{claim}(Reduction Formula 1)
Suppose $f: pd(S_\rvx)\rarrow Reals$.
Define

\beq
I(f) =
r(n, N_\rvx) \int_{pd(S_\rvx)} \calD P\;
\frac{
e^{n H(P)}
}{
\sqrt{ \prod \{P\}}
}
f(P)
\;,
\eeq
where

\beq
r(n, N_\rvx) = \left(
\frac{n}{2 \pi}
\right)^{\frac{1}{2}(N_\rvx -1)}
\;.
\eeq
Then

\beq
\sum_{\vec{x}} f(\pclass{x})
\approx I(f)
\;,
\label{eq:dec-formula-one}
\eeq
and

\beq
I(1)\approx  N_\rvx^n
\;.
\label{eq:I-one}\eeq
\end{claim}
{\bf proof:}

\begin{subequations}\label{eq:dec-1-proof}
\begin{eqnarray}
\sum_{\vec{x}} f(\pclass{x})
&=&
\sum_{\class{x} \in \calC(S_\rvx^n)} | \class{x} | f(\pclass{x})
\\
&=&
|\calC(S_\rvx^n)|
\frac{
\sum_{\class{x} \in \calC(S_\rvx^n)} | \class{x} | f(\pclass{x})
}{
\sum_{\class{x} \in \calC(S_\rvx^n)}
}
\\
&=&
|\calC(S_\rvx^n)|
\frac{
\sum_{\pclass{x} \in \calP(S_\rvx^n)} | \class{x} | f(\pclass{x})
}{
\sum_{\pclass{x} \in \calP(S_\rvx^n)}
}
\label{eq:c-to-p}
\\
&\approx&
|\calC(S_\rvx^n)|
\frac{
\int_{pd(S_\rvx)} \calD P \; | \class{x}| f(P)
}{
\int_{pd(S_\rvx)} \calD P \;
} \\
&\approx&
I(f)
\;.
\eeqa
In  Eq.(\ref{eq:dec-1-proof}),
we went from  line (d) to (e)
 by substituting previously derived values for
$|\calC(S^n_\rvx)|$, $|C(\vec{x})|$ and $\int_{pd(S_\rvx)} \calD P 1$.
This proves Eq.(\ref{eq:dec-formula-one}).

If we substitute $f=1$ into the LHS of
Eq.(\ref{eq:dec-formula-one}), we get $N^n_\rvx$.
But what if we substitute $f=1$ into the RHS of
Eq.(\ref{eq:dec-formula-one}) Does this also yield
$N^n_\rvx$? Yes.  Let's see how.
Define
$\Delta P(x) = P(x) - \frac{1}{N_\rvx}$.
If we expand $H(P)$ about the point $\Omega$ ,
we get:
(See Appendix \ref{app:taylor-exp} for a
compendium of Taylor expansions related to Information Theory)

\beq
H(P) \approx \ln N_\rvx - \frac{N_\rvx}{2}
\sum_x [\Delta P(x)]^2 + {\cal O}((\Delta P)^3)
\;.
\eeq
For large $n$, most of $I(1)$
comes from the vicinity of $\Omega$. Since $\Omega$
is far away from the boundary of the probability simplex,
the constraint $\theta(P\geq 0)$  can be ignored in
$I(1)$. Thus,
$I(1)$
can be approximated by:

\begin{subequations}
\label{eq:H-near-omega}
\begin{eqnarray}
I(1) &\approx&
 r(n, N_\rvx) N_\rvx^{n + \frac{N_\rvx}{2}}
\int \calD \Delta P \; \delta (\sum_x \Delta P(x))
\exp\left( -\frac{n N_\rvx}{2} \sum_x [\Delta P(x)]^2\right)\\
&\approx&
 N_\rvx^n
\;.
\eeqa
In Eq.(\ref{eq:H-near-omega}), to go from line (a) to (b),
we performed the integration
using the Gaussian integration formulae of Appendix \ref{app:gaussian-int}.
QED

\begin{claim}(Reduction Formula 2)
Suppose $f: pd(S_{\rvx, \rvy})\rarrow Reals$.
Define

\begin{eqnarray}
\lefteqn{
J(f)=\frac{
 r(n, N_{\rvx \rvy}-N_\rvy)
}{
[ \prod\{P_\rvy\}]^{\frac{1}{2}(N_\rvx-1)}
}
\int \calD P_{\rvx, \rvy}\;
}\nonumber \\
&&
\theta(P_{\rvx, \rvy}\geq 0)
\prodset{\delta(P_\rvy(y) - \pclass{y}(y))}{y}\nonumber\\
&&
\frac{
\exp[ n H_{P_{\rvx, \rvy}}(\rvx| \rvy)]
}{
\sqrt{\prodset {P(x|y)}{x,y}}
}
f(P_{\rvx, \rvy})
\;.
\end{eqnarray}
Then

\beq
 \sum_{\vec{x}} f(\pclasstwo{x}{y})\approx J(f)
\;,
\label{eq:dec-formula-two}
\eeq
and

\beq
J(1) \approx N_\rvx^n
\;.
\eeq
\end{claim}
{\bf proof:}
Clearly,

\beq
\sum_{\vec{x}} f(\pclasstwo{x}{y})
=
\sum_{\vec{x_1}, \; \vec{y_1}} \delta(\vec{y_1}, \vec{y}) f(\pclasstwo{x_1}{y_1})
\;.
\label{eq:j-sum-vecs}
\eeq
We would like to transform the sum over the words $\vec{x_1}$ and
$\vec{y_1}$ into a sum over ``coarser" items: namely, a sum over CC's like $\classtwo{x_1}{y_1}$.
These CC's are in 1-1 correspondence with
their probability distributions $\pclasstwo{x_1}{y_1}$,
and a sum over these distributions can be approximated by
an integral over the probability simplex $pd(S_{\rvx, \rvy})$.
All this can be accomplished if we approximate the Kronecker
delta for  points  $\vec{y}$ by a suitably normalized Dirac delta
function for distributions $\pclass{y}$. So let us do the following
replacement:

\beq
\delta(\vec{y_1}, \vec{y}) \rarrow K
\prodset{\delta( \pclass{y_1}(y) - \pclass{y}(y)}{y}
\;.
\label{eq:delta-coarsening}
\eeq
We choose the value of the normalization constant $K$  to be

\beq K = \frac{ \sqrt{\prod\{\pclass{y}\}} \exp[-n H(\pclass{y})]
}{ r(n, N_\rvy)\; \delta(\sum_y \pclass{y}(y) - 1) } \;.
\label{eq:k-choice} \eeq (Division by a Dirac delta function is allowed
as an intermediate step, before taking the $\epsilon$ parameter of
Eq.(\ref{eq:dirac-del}) to zero.) The reason for choosing this
value for $K$ is as follows. Using
 Reduction Formula 1 and Eq.(\ref{eq:delta-coarsening}),
one gets

\begin{eqnarray}
\lefteqn{
1 = \sum_{\vec{y}} \delta(\vec{y_1}, \vec{y})
\approx
r(n, N_\rvy) \int \calD P_\rvy \;
\theta(P_\rvy \geq 0)
\delta( \sum_y P_\rvy(y) - 1)
}
 \nonumber\\
&&
\frac{
\exp[ n H(P_\rvy)]
}{
\sqrt{\prod\{P_\rvy\}}
}
K
\prodset{\delta( \pclass{y_1}(y) - P_\rvy(y)}{y}
\;.
\end{eqnarray}
The previous equation is satisfied for the value of $K$ given
by Eq.(\ref{eq:k-choice}).

To show Eq.(\ref{eq:dec-formula-two}),
one replaces the Kronecker delta $\delta(\vec{y_1}, \vec{y})$
in the RHS of Eq.(\ref{eq:j-sum-vecs})
by a coarser delta, in accordance with  the
prescription Eq.(\ref{eq:delta-coarsening}).
Then one applies  Reduction Formula 1
to the result. This proves Eq.(\ref{eq:dec-formula-two}).

If we substitute $f=1$ into the LHS of
Eq.(\ref{eq:dec-formula-two}), we get $N^n_\rvx$. But what if we
substitute $f=1$ into the RHS of Eq.(\ref{eq:dec-formula-two})
Does this also yield $N^n_\rvx$? Yes. Here is a sketch of the
proof. The proof comprises two main steps: First, use the results
of Appendix \ref{app:int-fixed-margi} to convert $J(1)$ from an
integral of the form $\int \prodset{ dP(x,y)}{x,y}(\cdot)$ to a
product over all $y$ of integrals of the form $\int \prodset{
dP(x|y)}{x}(\cdot)$. Second, apply the Gaussian integration
formulae of Appendix \ref{app:gaussian-int}. QED

\section{Noiseless Coding}
In this section we will discuss
Noiseless Coding (i.e., a coding used in compression).
In particular, we will calculate the probability of
error, in the limit of  large word size $n$,
for compression using the Csisz\'ar-K\"orner (CK) universal code.

\subsection{Error Model}
This section reviews the usual error model for
compression using CK universal code. Subsequent sections
will apply CCRG to it.

\begin{figure}[h]
    \begin{center}
    \epsfig{file=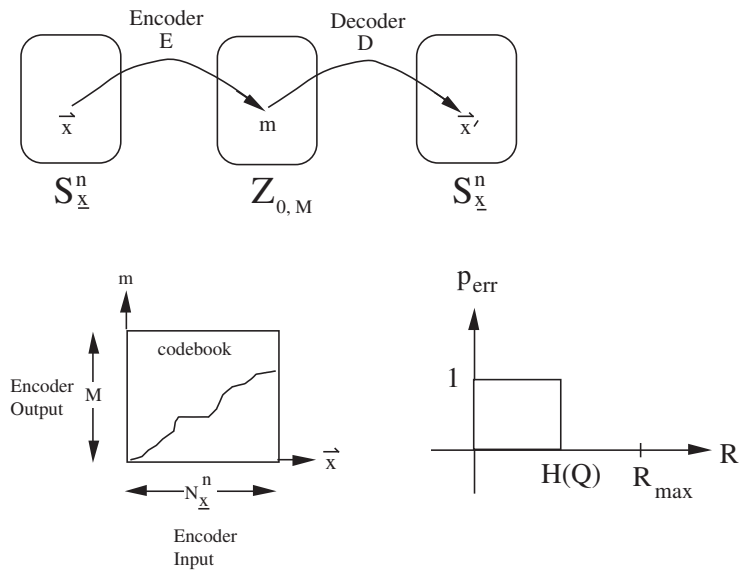, height=3.0in}
    \caption{Encoding and Decoding maps for Noiseless Coding.}
    \label{fig:nless-maps}
    \end{center}
\end{figure}

A {\it block source} emits a stream  of $n$-letter words
$(x^1, x^2, \ldots, x^n) = x^{/n} = \vec{x}\in S^n_\rvx$.
Each word is modelled as a sequence of
$n$ i.i.d. random variables $\rvx^i$ distributed
according to $Q(x^i)$, where $i\in Z_{1, n}$.

Suppose that, as shown in Fig.\ref{fig:nless-maps}:
(1)Each word $\vec{x}\in S^n_\rvx$ is mapped by an
{\it encoder function} $E$ into a message $E(\vec{x}) = m\in Z_{0, M}$. (2)
Each message $m$ is in turn mapped by a {\it decoder function}
$D$ into a word
$D(m) = \vec{x'}\in S^n_\rvx$.
A {\it block code} is characterized by: the probability distribution $Q$
of its source,
its encoder function $E$ and its decoder function $D$.
The block code is said to be {\it universal} if $E$ and $D$ do not depend on $Q$.

Assume that $|{\rm Image} (E)| = |E(S^n_\rvx)|\approx M$.
The {\it compression factor} or {\it code rate} $R$
of the encoder is defined by

\beq
R = \frac{\ln M}{n} = (\ln 2) \frac{\log_2 M}{n}
\;.
\eeq
Note that if $N_\rvx = 2$, then
$\frac{\log_2 M}{n} = \frac{n_{out}}{n_{in}}$
where $n_{out}$ (ditto, $n_{in}$ )
 is the encoder output (ditto, input) measured in bits.
Note that $R\leq \ln N_\rvx$ because $M\leq N^n_\rvx$.
For a {\it fixed rate} block code, $R$ is fixed as $n\rarrow \infty$

The {\it probability of error} for the code is given by

\beq
p_{err} = \sum_{\vec{x}} Q(\vec{x}) \theta (D\circ E(\vec{x}) \neq \vec{x})
\;.
\eeq

Assume a fixed rate block code and let

\beq
\RV = R - N_\rvx \frac{\ln (n+1)}{n}
\;.
\eeq
Of course, for large $n$ , $\RV \approx R$.
Let

\beq
A_{pass} = \{ \vec{x} \in S_\rvx^n : \; H(\pclass{x}) \leq \RV \},
\; A_{stop} = S_\rvx^n - A_{pass}
\;.
\eeq
$|A_{pass}| \leq M$  because
\beqa
|A_{pass}| &=& \sum_{\class{x}}
| \class{x}| \theta( H(\pclass{x}) \leq \RV)
\\
&\leq&  \sum_{\class{x}} e^{n H(\pclass{x})} \theta( H(\pclass{x}) \leq \RV)
\\
&\leq& e^{n\RV} | \calC(S^n_\rvx)|\\
&\leq& e^{n\RV} (n+1)^{N_\rvx} = e^{n R} =  M
\;.
\eeqa
If $|A_{pass}|\approx M$, then $|A_{stop}| \approx
N_\rvx^n - M  = e^{n \ln N_\rvx} - e^{nR}$. Since
$R\leq \ln N_\rvx$, $|A_{stop}| >> |A_{pass}|$ for large $n$.

We can number
the elements of $A_{pass}$ from 1 to $|A_{pass}|$. Call
$m(\vec{x})$ the number assigned to  $\vec{x}\in A_{pass}$.
The {\it CK universal code} is a fixed rate
block code with  encoding and decoding functions
defined by:

\beq
E(\vec{x}) =
\left\{
\begin{array}{l}
m(\vec{x}) {\rm \; if\;} \vec{x}\in A_{pass}\\
0\;{\rm if}\; \vec{x}\notin A_{pass}
\end{array}
\right.
\;,
\eeq

\beq
D(m) =
\left\{
\begin{array}{l}
E^{-1}(m)\; {\rm if}\; m\in Z_{1, |A_{pass}|}\\
{\rm any}\; \vec{x}\notin A_{pass}\; {\rm if }\; m=0
\end{array}
\right.
\;.
\eeq
Note that low entropy  words  (i.e., those $\vec{x}$ with
$H( \pclass{x})<R$)  belong to $A_{pass}$ and are
coded, whereas the high entropy words
(i.e., those $\vec{x}$ with
$H( \pclass{x})>R$) belong to $A_{stop}$ and
are not coded. Thus, the CK universal code can be described
as a low pass filter of word entropy.
Why are low entropy words preferable to high entropy ones for coding?
Because for $R=H(Q)$,
$Q(A_{pass})$ and $Q(A_{stop})$ are comparable even though
$|A_{pass}|<< |A_{stop}|$.
Note that

\beq
\theta( D\circ E(\vec{x}) \neq \vec{x} ) =
\theta(\vec{x} \notin A_{pass}) =
\theta( H(\pclass{x})> R)
\;.
\eeq
Thus, for CK universal coding,

\beq
p_{err} = \sum_{\vec{x}} Q(\vec{x}) \theta( H(\pclass{x})> R)
\;.
\eeq
Applying Reduction Formula 1 to the RHS of the previous equation yields

\beq
p_{err} \approx
r(n, N_\rvx) \int_{pd(S_{\rvx})} \calD P \;
\frac{e^{-n D(P//Q)}
}{
\sqrt{\prod\{P\}}
}
\theta( H(P) > R)
\;.
\label{eq:perr-nless}
\eeq
In the previous equation,
the exponential inside the integral reaches its maximum value when $D(P//Q)=0$.
If we approximate $P$ by $Q$ in the theta function
of the integrand, then we can pull the theta function
out of the integral. Doing this yields

\beq
p_{err} \approx \theta( H(Q) > R)
\;.
\label{eq:perr-nless-crude}
\eeq
In other words, if the compression factor $R$ is
larger (ditto, smaller) than $H(Q)$,
then the probability of error is zero (ditto, one).
The next few sections of this paper will be dedicated
to improving this estimate of $p_{err}$.
\begin{figure}[h]
    \begin{center}
    \epsfig{file=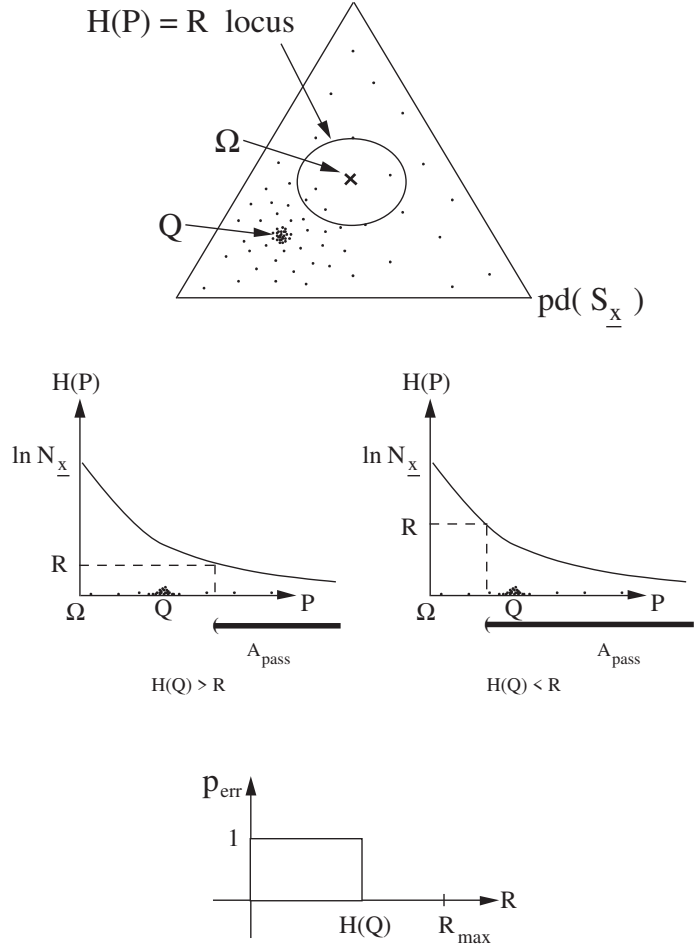, height=5.0in}
    \caption{$A_{pass}$ when $H(Q)$
    is greater or smaller than $R$.
Strictly speaking, $A_{pass}$ is a set of $\vec{x}$,
and what we are showing is ${\cal P}(A_{pass})$ instead of
$A_{pass}$.}
    \label{fig:nless-plots}
    \end{center}
\end{figure}

\subsection{Old Approximation for  $p_{err}$}\label{sec:old-nless-perr}

In this section, we will review the standard
calculation (see \cite{blahut}) of the error exponent for CK universal coding.
In the next section, we will calculate the
error exponent (and much more) using CCRG.

The standard way of finding the error exponent for
CK universal coding is
equivalent to using Laplace's Method
to find the leading term of an
asymptotic expansion of Eq.(\ref{eq:perr-nless}).
To apply Laplace's Method, we must
minimize $D(P//Q)$ over all $P\in pd(S_\rvx)$,
subject to the inequality constraint $H(P)>R$.

To obtain a minimum point $\vec{x}^*\in Reals^n$ of a smooth,
real-valued function $f(\vec{x})$, subject to equality constraints
$c_j(\vec{x})=0$ for $j\in C_{eq}$, one can use the well known
method of Lagrange multipliers. But suppose that, in addition to
these equality constraints, $\vec{x}^*$ must also satisfy
inequality constraints $c_j(\vec{x})\geq 0$ for $j\in C_{geq}$. To
obtain a minimum $\vec{x}^*$ in this more complicated case, one
can generalize the method of Lagrange multipliers. Kuhn and
Tucker, among others, have done this. Let $J = C_{eq} \cup
C_{geq}$, and define the Lagrangian function $\calL= f(\vec{x}) -
\sum_{j\in J}\lam_ j c_j(\vec{x})$. According to Kuhn-Tucker, the
minimum point $\vec{x}$ and the Lagrange multipliers
$(\lam_j)_{\forall j\in J}$ must satisfy the {\it Kuhn-Tucker
conditions}\cite{Fletcher} given by (1) $\nabla_{\vec{x}} L = 0$,
(2) $\forall j\in C_{eq}$, one has $c_j(\vec{x}) = 0$ (3)$\forall
j\in C_{geq}$, one has $c_j(\vec{x})\geq 0$, $\lam_j \geq 0$ and
$\lam_j c_j(\vec{x}) = 0$.

Let
\beq
\calL = D(P//Q) - \lam(H(P) -R) + \mu(\sum_x P(x) -1)
\;.\eeq
For the problem we are considering here, the Kuhn-Tucker conditions are
(1) $\forall x,\;  \pder{\calL}{P(x)} = 0$,
(2)$\sum_x P(x) =1$,
(3) $H(P) -R \geq 0$, $\lam\geq 0$,  $\lam( H(P) - R) =0$.
We will assume that the inequality constraint is ``active" \cite{Fletcher}, in
which case condition (3) reduces to $H(P) = R$. Condition (1) implies:

\beqa
0 &=& \ln \frac{P(x)}{Q(x)} + 1 - \lam(-\ln P(x) - 1) + \mu\\
&=& \ln( P^{1+\lam}(x) ) - \ln Q(x) + 1 +\lam + \mu
\;.
\eeqa
The previous equation is satisfied by

\beq
P^\lamfun(x) = \frac{Q(x)^{\frac{1}{1+\lam}}}{Z}
\;,
\label{eq:pstar-nless}
\eeq
where

\beq
Z = \sum_x Q(x)^{\frac{1}{1+\lam}}
\;.
\eeq
This value for $P^\lamfun$ satisfies
$\sum_x P^\lamfun(x)=1$,
but does not yet satisfy $H(P^\lamfun)=R$.
The equation $H(P^\lamfun)=R$
defines a unique value of $\lam$.

Define

\beq
\gamma(\lam) = \min_{P, \mu} \calL = D(P^\lamfun//Q)
\;.
\eeq
Substituting the value for $P^\lamfun$
given by Eq.(\ref{eq:pstar-nless})
into $D(P^\lamfun//Q) $
 yields:

\beq
\gamma(\lam) = \lam R - (1+\lam)\ln Z
\;.
\label{eq:gamma-lam}
\eeq
$P^\lamfun$ and $\gamma(\lam)$  still depend on a
parameter $\lam$ which is specified implicitly by
the equation $H(P^\lamfun) = R$.
In fact, one can show that
$H(P^\lamfun) = R$ iff $\frac{d\gamma(\lam)}{d\lambda}=0$.

Define
the {\it error exponent} $\gamma$ by

\beq
\gamma = \max_{\lam\geq 0}\gamma(\lam)
\;.
\eeq
It is now clear that $p_{err}$ given by Eq.(\ref{eq:perr-nless})
can be approximated by:

\beq
p_{err} \approx e^{-n \gamma}, \;\;{\rm where}\;\;
\gamma = \max_{\lam\geq 0} [\lam R - (1+\lam)Z(\lam)]
\;.
\label{eq:perr-trad}
\eeq
Eq.(\ref{eq:perr-trad}) is
the traditional \cite{blahut} asymptotic approximation
for the probability of error for CK universal coding.

\subsection{New (CCRG) Approximation for $p_{err}$}\label{sec:new-nless-perr}

In this section and the next one,
we will use CCRG to calculate the probability of
error for compression using
the CK universal code.
In this section, we will calculate  $p_{err}$
as given by Eq.(\ref{eq:perr-nless}), assuming
that we have rescaled the variables of the RHS of
Eq.(\ref{eq:perr-nless}) so
that the integrand is a Gaussian. In the next section,
we will derive the RG equations that characterize this
rescaling.

Let $P = P - Q$,
$\Delta H(P) = H(P) -H(Q)$, and
$\Delta R = R - H(Q)$.
Hence,
$\theta( H(P) > R) = \theta( \Delta H(P) > \Delta R)$.

Let
\beq
\calL = D(P//Q) - \lam (\Delta H(Q) - \Delta R)
+ \mu (\sum_x P(x) -1)
\;.
\label{eq:lagrangian-nless-exact}
\eeq
Minimizing this Lagrangian with respect to $P, \lam, \mu$
gives the saddle (or boundary) point $P^*$ that dominates the integral given by
Eq.(\ref{eq:perr-nless}). Unfortunately,
finding an explicit expression for $P^*$
is not possible.

Define
{\it test fractions} $\Phi_0$ and $\Phi_1$ by

\beq
\Phi_0 =
\left|
\frac{ D(P^*//Q)}
{ \sum_x \frac{[\Delta P^*(x)]^2}{2Q(x)} }
-1
\right|
\;,
\eeq

\beq
\Phi_1 =
\left|
\frac{ H(P^*) - H(Q)}
{- \sum_x \Delta P^*(x) \ln Q(x) }
-1
\right|
\;.
\eeq
$\Phi_0$ (ditto, $\Phi_1$) measures how
much $D(P^*//Q)$ (ditto, $\Delta H(P^*)$)
differs from the leading term of its Taylor expansion
about $Q$. (See Appendix \ref{app:taylor-exp} for a
compendium of Taylor expansions related to Information Theory).

Suppose we have rescaled the variables in
the RHS of Eq.(\ref{eq:perr-nless})
so that after rescaling,
we are in the ``Gaussian region": $\Phi_0 << 1$ and $\Phi_1 <<1$.
Then Eq.(\ref{eq:perr-nless}) can be
approximated by

\begin{eqnarray}
\lefteqn{
p_{err} \approx
\frac{
r(n, N_\rvx)
}{
\sqrt{ \prod \{ P^*\} }
}
\int \calD \Delta P\;
\delta(\sum_x \Delta P(x))
}\nonumber \\
&&
\exp[ -n \sum_x \frac{[\Delta P(x)]^2}{2 Q(x)}]
\;\;
\theta(-\sum_x \Delta P(x) \ln Q(x) > \Delta R)
\;.
\label{eq:perr-nless-quad}
\end{eqnarray}
(For large $n$, if $Q$
is not too close to the boundary of the probability simplex,
then the constraint $\theta(P\geq 0)$  can be ignored.)

In the Gaussian region, we  can also approximate
Eq.(\ref{eq:lagrangian-nless-exact}) by

\beq
\calL = \sum_x \frac{[\Delta P(x)]^2}{2 Q(x)}
- \lam (-\sum_x \Delta P(x)\ln Q(x) - \Delta R)
+ \mu (\sum_x P(x) -1)
\;.
\label{eq:lagrangian-nless-quad}
\eeq
Minimizing this Lagrangian with respect to $P, \lam, \mu$
gives the  point $P^*$ that dominates the integral given by
Eq.(\ref{eq:perr-nless-quad}). Finding an explicit expression for
$P^*$ in the Gaussian region is possible. $\pder{\calL}{P(x)}=0$ gives:

\beq \frac{\Delta P(x)}{Q(x)} + \lam \ln Q(x) + \mu =0 \;. \eeq
Enforcing the constraints $-\sum_x \Delta P(x)\ln Q(x) = \Delta R$
and $\sum_x P(x) =1$ then yields

\beq
\Delta P^*(x) =  B(x)\Delta R
\;,
\label{eq:pstar-nless-quad}
\eeq
where

\beq
B(x) = \frac{\beta(x)Q(x)}{\av{\beta^2}}
\;,
\eeq

\beq
\beta(x) = -[ \ln Q(x) + H(Q)]
\;,
\eeq

\beq
\av{\beta} = \sum_x Q(x) \beta(x) = 0
\;,
\eeq

\beq
\av{\beta^2} = \sum_x Q(x) \beta^2(x)
\;.
\eeq

On the RHS of Eq.(\ref{eq:perr-nless-quad}),
we can apply the Gaussian Integration Formulae of Appendix \ref{app:gaussian-int}. We can also
substitute there the value for $P^*$
given by Eq.(\ref{eq:pstar-nless-quad}).
Doing so finally
gives

\beq
p_{err} \approx
\frac{1}{2 u} \erfc
\left(
\Delta R
\sqrt{\frac{n}{2 \av{\beta^2}}}
\right)
\;,
\label{eq:perr-erfc-nless}
\eeq
where

\beq
u = \sqrt{
\prodset{ 1 + \frac{\beta(x)\Delta R }{\av{\beta^2}} }{x}
}
\;.
\eeq
Appendix \ref{app:err-fun} reviews  some basic
properties of the Error Function erf() and its
complement erfc().

\subsection{RG Equations}\label{sec:new-nless-rg-eqs}
In this section, we will calculate the RG equations for
compression using CK
universal coding.

Important: In this section, $\Delta P^\sfun$ describes the motion,
upon successive rescalings, of the  point that
dominates the integral of Eq.(\ref{eq:perr-nless}).

Consider the argument of the exponential
in the integrand of Eq.(\ref{eq:perr-nless}).
It should be invariant under a change of scale:

\beq
n^\wedge D^\wedge(P//Q) = n D(P//Q)
\;.
\eeq
If for some $\delta s$ such that $0\leq\delta s<<1$,

\beq
n^\wedge = e^{\delta s} n
\;,
\eeq
then

\beq
D^\wedge(P//Q)  = D(P^\wedge//Q) =e^{-\delta s} D(P//Q)
\;.
\label{eq:scaled-dist}
\eeq
Define
\beq
P^\wedge(x)= P^{(\delta s)} (x) =
(1-\gamma_0 \delta s) P(x) + (\gamma_0 \delta s )Q(x)
\;.
\eeq
Then, for $s>0$,

\beq
\pder{ \Delta P^\sfun(x)}{s}
=
-\gamma_0(P^\sfun, Q)
\Delta P^\sfun(x)
\;,
\eeq
where

\beq
\gamma_0(P, Q) =
\lim_{s\rarrow 0}
\frac{
(-1)  \pder{\Delta P^\sfun}{s}
}{
\Delta P^\sfun
}
\;.
\eeq
By virtue of Eq.(\ref{eq:scaled-dist}),

\beq
\pder{ D(P^\sfun//Q)}{s}
=
-D(P^\sfun//Q)
\;,
\eeq
where

\beq
1  =
\lim_{s\rarrow 0}
\frac{
(-1)  \pder{D(P^\sfun//Q)
}{s}
}{
D(P^\sfun//Q)
}
\;.
\eeq
Note that

\beqa
\lim_{s\rarrow 0}
\pder{ D(P^\sfun//Q)}{s}
&=&
\lim_{s\rarrow 0}
\sum_x
\pder{P^\sfun}{s} (\ln \frac{P^\sfun(x)}{Q(x)} + 1)\\
&=& -\gamma_0 \sum_x
\Delta P(x) (\ln \frac{P(x)}{Q(x)} + 1)
\\
&=&
-\gamma_0 [ D(P//Q) + D(Q//P)]
\;.
\eeqa
Thus

\beq
\gamma_0(P, Q) =
\frac{
D(P//Q) }{
D(P//Q) + D(Q//P)
}
\;.
\eeq

Now consider the theta function
in the integrand of Eq.(\ref{eq:perr-nless}).
It too should be invariant under a change of scale:

\beq
\theta( \Delta H^\wedge(P) > \Delta R^\wedge) =
\theta( \Delta H(P) > \Delta R)
\;.
\eeq
If for some $\delta s$ such that $0 \leq \delta s << 1$,

\beq
\Delta R^\wedge = e^{-\gamma_1 \delta s}\Delta R
\;,
\label{eq:scaled-del-R}
\eeq
then

\beq
\Delta H^\wedge(P) =
\Delta H(P^\wedge) =
e^{-\gamma_1 \delta s}\Delta H(P)
\;.
\label{eq:scaled-del-H}
\eeq
Eqs. (\ref{eq:scaled-del-R})
 and (\ref{eq:scaled-del-H}) imply

\beq
\pder{ V^\sfun}{s} = -\gamma_1( P^\sfun, Q) V^\sfun
\;,
\eeq
where

\beq
V^\sfun = \left(
\begin{array}{c}
\Delta R^\sfun \\
\Delta H (P^\sfun)
\end{array}
\right )
\;,
\eeq
and

\beq
\gamma_1(P, Q) =
\lim_{s\rarrow 0}
\frac{
(-1)  \pder{\Delta R^\sfun}{s}
}{
\Delta R^\sfun
}
=
\lim_{s\rarrow 0}
\frac{
(-1)  \pder{\Delta H(P^\sfun)}{s}
}{
\Delta H(P^\sfun)
}
\;.
\eeq
Note that

\beq
\Delta H(P) = -\sum_x \Delta P(x) \ln P(x) + D(Q//P)
\;.
\eeq
Interchanging $P$ and $Q$ in the previous equation also yields:

\beq
-\Delta H(P) = +\sum_x \Delta P(x) \ln Q(x) + D(P//Q)
\;.
\eeq
Note that

\beqa
\lim_{s\rarrow 0}
\pder{\Delta H(P^\sfun)}{s}&=&
(-1)  \lim_{s\rarrow 0}
\sum_x
\pder{P^\sfun}{s} (\ln P^\sfun(x) + 1)\\
&=&
\gamma_0 [ -\Delta H(P) + D(Q//P)]
\;.
\eeqa
Thus,

\beq
\gamma_1(P, Q) =  \left(
1 -
\frac{D(Q//P)}{\Delta H(P)}
\right)
\gamma_0(P, Q)
\;.
\eeq

We will call $\gamma_0$ and $\gamma_1$ the {\it critical exponents}
for $\Delta P^\sfun$ and $\Delta R^\sfun$, respectively.

Note that $\gamma_0(P, Q)$ and $\gamma_1(P, Q)$
both tend to $\frac{1}{2}$ as $P\rarrow Q$.
Note  also that
$\gamma_0$ and $\gamma_1$ are related to the test fraction $\Phi_1$
as follows. Define $\phi\geq 0$ by

\beq
\phi =
\left|
\frac{\gamma_1}{\gamma_0} - 1
\right|
=\left|
\frac{
D(Q//P)}{\Delta H(P)}
\right|
\;.
\eeq
Then

\beqa
\Phi_1
&= &
\left|
\frac{
\Delta H(P) + \sum_x \Delta P(x) \ln Q(x)
}{
\sum_x \Delta P(x) \ln Q(x)
}
\right|
\\
&=&
\left |
\frac{ D(P//Q)}{ \Delta H(P) + D(P// Q) }
\right |\\
&\approx &
\left |
\frac{\phi}{1+\phi}
\right |
\;.
\eeqa

In conclusion, we must solve the following pair of
coupled RG equations,

\beq
\pder{ \Delta P^\sfun(x)}{s}
=
-\gamma_0(Q + \Delta P^\sfun, Q)
\Delta P^\sfun(x)
\;
\eeq
for all $x\in S_\rvx$, and

\beq
\pder{ \Delta R^\sfun}{s} = -\gamma_1( Q + \Delta P^\sfun, Q) \Delta R^\sfun
\;.
\eeq
We must solve this pair of RG equations subject to the
following pair of boundary conditions:
At $s=0$:
\beq
\Delta R^{(0)} = \Delta R
\;,
\eeq
and
at  $s= \sfin$:

\beq
\Delta P^{ (\sfin)} (x) =
B(x)\Delta R^{(\sfin)}
\;.
\eeq
 $\sfin$ is defined as any $s$ large enough for the following
to be true: $\Phi_0(P^{(\sfin)}, Q)<<1$ and $\Phi_1(P^{(\sfin)}, Q)<<1$.

Section \ref{sec:comp-results} describes  a computer program
  called WimpyRG-C1.0 that
solves these RG equations.

\section{Noisy Coding}
In this section, we will discuss Noisy Coding
(i.e., a coding used in channel transmission). In particular,
we will calculate the probability of error,
in the limit of  large word size $n$,
for channel transmission using random encoding and
maximum-likelihood  decoding.

\subsection{Error Model}
In this section we will review the error model for
channel transmission using random encoding and
maximum-likelihood decoding. Subsequent sections will apply CCRG to it.

\begin{figure}[h]
    \begin{center}
    \epsfig{file=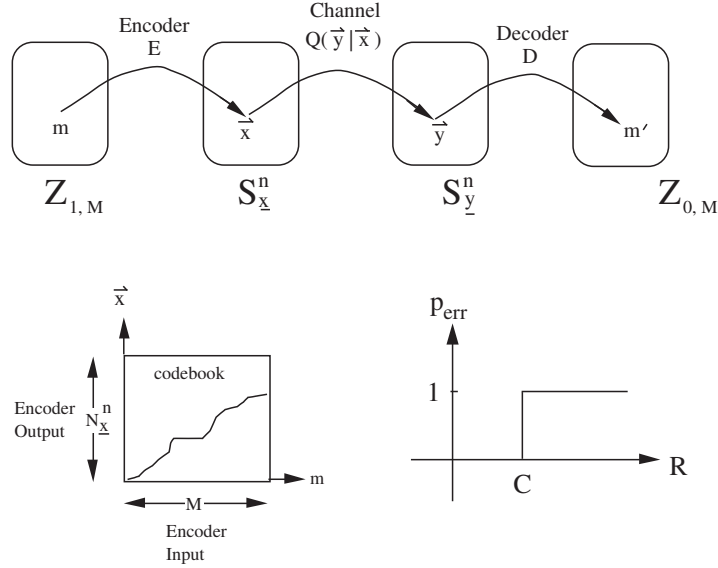, height=3.0in}
    \caption{Encoding, Channel and Decoding
    maps for Noisy Coding.}
    \label{fig:noisy-maps}
    \end{center}
\end{figure}

Suppose that, as shown in Fig.\ref{fig:noisy-maps}:
(1)Each message $m\in Z_{1, M}$
is mapped by an {\it encoder function} $E$ into a word
$\vec{x}\in S^n_\rvx$.
(2) A {\it channel} $Q(\vec{y}|\vec{x})$
gives the probability that word $\vec{x}\in S^n_\rvx$
is mapped into word $\vec{y}\in S^n_\rvy$.
(3)Each word $\vec{y}$ is then mapped by a
{\it decoder function} $D$ into message $m'\in Z_{0, M}$.
We assume a {\it discrete memoryless channel}, by which we
mean that

\beq
Q(\vec{y}| \vec{x}) = \prodset{Q(y^i | x^i)}{i\in Z_{1, n}}\;
\;.
\label{eq:q-vecy-at-vecx}
\eeq
An $(M, n)$ {\it channel code} is characterized by its
encoding function $E$,
the conditional probability of its channel $Q(\vec{y} | \vec{x})$,
and its decoding function $D$.

Let $p_{err|m}$ be the probability of error when message $m\in Z_{1,M}$ exits
the encoder. Then

\beq
p_{err | m} = Pr\{ D(\vec{\rvy})\neq m | \vec{\rvx} = E(m) \}
\;.
\label{eq:perr-at-m-noisy}
\eeq

The {\it code rate} $R$ of the encoder is defined by

\beq
R = \frac{\ln M}{n} = ( \ln 2)\frac{\log_2 M}{n}
\;.
\eeq
Note that if $N_\rvx = 2$,
then $\frac{\log_2 M}{n} = \frac{n_{in}}{n_{out}}$
(careful: for noiseless coding $\frac{\log_2 M}{n} = \frac{n_{out}}{n_{in}}$
instead), where $n_{out}$ (ditto, $n_{in}$) is the
encoder output (ditto, input) measured in bits.

The {\it maximum achievable rate} $R_{max. ach.}$ is defined by:

\beq
R_{max. ach.} =
\lim_{\epsilon \rarrow 0}
\lim_{M\rarrow \infty} \sup
\{
\frac{\ln M}{n} : \;
\exists (n, E, D) \forall m, p_{err | m}(n, E, D) < \epsilon
\}
\;.
\eeq

The {\it information capacity C} is defined by:

\beq
C = \max_{Q_\rvx \in pd(S_\rvx)} H_{Q_{\rvx, \rvy}}(\rvx : \rvy)
\;.
\eeq
The fact that $R_{max. ach.}= C$
is essentially Shannon's Noisy Coding (or  ``Second")
Theorem).

Eq.(\ref{eq:perr-at-m-noisy})
can be re-expressed as

\beqa
p_{err | m}
&=&
\sum_{\vec{y}}
Pr\{ D(\vec{\rvy}) \neq m | \vec{\rvx} = E(m), \vec{\rvy} = \vec{y} \}
Pr\{ \vec{\rvy} = \vec{y} | \vec{\rvx} = E(m) \}
\\
&=&
\sum_{\vec{y}} \theta(D(\vec{y}) \neq m) Q(\vec{y} | \vec{x}(m))
\\
&=& 1-
\sum_{\vec{y}} \theta(D(\vec{y}) = m) Q(\vec{y} | \vec{x}(m))
\label{eq:perr-at-m-2}
\;.
\eeqa

A {\it random encoder} $E$ is defined
by choosing each component of $\vec{x} = E(m)$
independently from the other components and according
to the probability distribution $Q(x)$. With such an encoder,

\beqa
\lefteqn {p_{err} =
\sum_{m\in Z_{1, M}, E} p_{err | m, E} P(E) P(m)}\\
&=& 1- \sum_{m \in Z_{1, M}} P(m)
\prodset{
 \sum_{\vec{x}(m_1)\in S_\rvx^n}
 Q(\vec{x}(m_1))}
{m_1\in Z_{1, M}}
\sum_{\vec{y}}
\theta( D(\vec{y})= m)
Q(\vec{y} | \vec{x}(m))
\;.\nonumber\\
&&
\label{eq:perr-before-v}
\eeqa

Suppose  $\Gamma: S^n_\rvy \times Z_{1, M} \rarrow \{true, false\}$
is a condition, and $Good(\Gamma)$ is the set of all $\vec{y}$
for which there is a unique $m\in Z_{1, M}$ that satisfies
$\Gamma(\vec{y}, m) = true$.
Also let $Bad(\Gamma) = S^n_\rvy - Good(\Gamma)$.
One can define the decoding function $D$ implicitly in terms
of the condition $\Gamma$
as follows:

\beq
D(\vec{y}) =
\left\{
\begin{array}{l}
{\rm unique}\:
m \; {\rm such\;that}\; \Gamma(\vec{y}, m) = true,
\; {\rm if} \; \vec{y}\in Good(\Gamma)\\
0
 \; {\rm if} \; \vec{y}\in Bad(\Gamma)
\end{array}
\right.
\;.
\eeq
Hence, for $m\in Z_{1,M}$ and $\vec{y} \in Good(\Gamma)$,

\beq
\theta( D(\vec{y}) = m) =
\theta(\Gamma(\vec{y}, m) )
\;.
\eeq
The {\it maximum likelihood (ML) decoder} is defined
by the condition

\beq
\Gamma(\vec{y}, m) =
\left(
\frac{
Q(\vec{y} | \vec{x}(m))
}{
Q(\vec{y} | \vec{x}(m'))
}
> 1 \;
 \forall m' \in Z_{1,M}, m'\neq m
\right)
\;.
\label{eq:ml-discriminant}
\eeq
(As illustrated in Fig.\ref{fig:max-likelihood}, we
assume that $Bad(\Gamma)$ is negligibly small,
in the sense that, for all $m\in Z_{1, M}$ , $\sum_{\vec{y} \in Bad(\Gamma)}
Q(\vec{y}|\vec{x}(m))<<1$.)
Actually, the ML decoder is not optimal.
It can be shown\cite{blahut} that the optimal decoder is
one for which

\beq
\Gamma(\vec{y}, m) =
\left(
\frac{
Q(\vec{x}(m) |\vec{y} )
}{
Q(\vec{x}(m') |\vec{y} )
}
> 1 \;
\forall m' \in Z_{1,M}, m'\neq m
\right)
\;.
\eeq

\begin{figure}[h]
    \begin{center}
    \epsfig{file=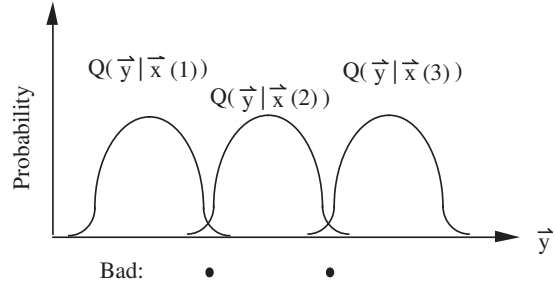, height=1.5in}
    \caption{Intuitive picture of condition Eq.(\ref{eq:ml-discriminant})
for Maximum Likelihood
decoder.}
    \label{fig:max-likelihood}
    \end{center}
\end{figure}

For each $(\vec{x}, \vec{y}) \in S^n_\rvx \times S^n_\rvy$,
define functions $v$ and $f$ by

\beq
v
(\vec{x}, \vec{y}) =
\sum_{\vec{x'}\in S_\rvx^n }
\theta
\left(
\frac{
Q(\vec{y} | \vec{x})
}{
Q(\vec{y} | \vec{x'})
}
>1
\right)
Q(\vec{x'})
\;,
\eeq
and

\beq
 f = 1 - v
\;.
\eeq
(mnemonic: $v$ stands for victory and $f$ for failure).

If we substitute into Eq.(\ref{eq:perr-before-v})
the value of $\theta( D(\vec{y}) = m)$ for ML decoding,
one finds for random encoding and ML decoding:

\beq
p_{err} =
1 - \sum_{\vec{x}, \vec{y}}
Q(\vec{y}|\vec{x}) Q(\vec{x}) [v(\vec{x}, \vec{y})]^{M-1}
\;.
\label{eq:perr-noisy-1}
\eeq
Later on, we will show that
$f\approx e^{-nC}$.
Since
$M = e^{nR}$,
it follows that for random encoding and ML decoding

\begin{subequations}
\label{eq:perr-noisy-crude}
\begin{eqnarray}
p_{err} &\approx & 1 -
(1- e^{-n C})^M \\
&\approx & 1 - \exp(-M e^{-nC}) \\
&\approx & 1 - \exp[-e^{n(R-C)}]\\
&\approx & 1 - \theta(R-C\leq 0) = \theta( R> C)
\label{eq:perr-noisy-crude-fin}
\;.
\eeqa
In Eq.(\ref{eq:perr-noisy-crude}),
we went from line (c)  to (d) by using the following easy to
prove identity: For all $x\neq 0$,

\beq
\theta(x >0) = \lim_{n\rarrow 0} \exp[-\exp(-nx) ]
\;.
\label{eq:theta-exp-exp}
\eeq
According to Eq.(\ref{eq:perr-noisy-crude-fin}),
if the code rate $R$ is larger (ditto, smaller)
than the channel capacity $C$, then the probability
of error is one (ditto, zero). The next few sections
of this paper will be dedicated to improving
this estimate of $p_{err}$.

\subsection{New (CCRG) Approximation for $p_{err}$}\label{sec:new-noisy-perr}
In this section and the next one, we will use
CCRG to calculate the probability of error for channel
transmission using random encoding and ML decoding.
This section will calculate $p_{err}$ as given by
Eq.(\ref{eq:perr-noisy-1}), assuming that we have
rescaled the variables on the RHS of Eq.(\ref{eq:perr-noisy-1})
so that the integrand is Gaussian. The next section will calculate
the RG equations that characterize this rescaling.

In what follows, we will use $Q(x, y)$ to mean
$Q(x, y) = Q(y|x) Q(x)$, where $Q(y|x)$ (ditto, $Q(x)$)
is the probability distribution that specifies the
transmission channel (ditto, the random encoding).
We will also use the following abbreviations:

\beq
C_1 = \sum_{x, y} Q(x, y) \ln \left (\frac{Q(x, y)}{Q(x) Q(y)}\right)
 =H_Q(\rvx : \rvy)
\;,
\eeq

\beq
\Delta R = R -C_1
\;,
\eeq

\beq
\Delta P(x, y) = P(x,y) - Q(x,y),
\;\;
\Delta P(x| y) = P(x|y) - Q(x|y)
\;,
\eeq

\beq
L_{xy} = \ln
\left(
\frac{ Q(x, y) }{Q(x) Q(y) }
\right)
\;.
\eeq
Note that $C_1$ is not equal to
the channel capacity $C$, but
$C = \max_{Q_\rvx\in pd(S_\rvx)}C_1$.

Applying Reduction Formula 1 to Eq.(\ref{eq:perr-noisy-1})
yields

\begin{subequations}
\label{eq:perr-v-M}
\begin{eqnarray}
p_{err} &=& 1 - \sum_{\vec{x}, \vec{y}}
Q(\vec{x}, \vec{y}) v^M\\
&\approx&
1 - r(n, N_{\rvx, \rvy})
\int_{pd(S_{\rvx,\rvy})} \calD P_{\rvx, \rvy}\;
\frac{\exp[-n D(P_{\rvx,\rvy} // Q_{\rvx, \rvy})]}
{
\sqrt{\prod \{ P_{\rvx,\rvy} \}}
}\;\;
v^M
\;.
\eeqa
For $n>>1$, and fixed $R$, $M= e^{nR}>>1$.
Later on we will show that $0\leq f <<1$.
The inequalities $M>>1$, and  $0\leq f <<1$, and
Eq.(\ref{eq:theta-exp-exp}) imply

\beq
v^M = (1-f)^M \approx e^{-Mf} =
e^{- \exp(n R + \ln f)} \approx \theta(R + \frac{\ln f}{n}<0)
\;.
\label{eq:v-M-theta}
\eeq

Our next goal is to calculate $\ln(f)$.
One has

\beq
f (\vec{x}, \vec{y}) =
\sum_{\vec{x'}, \vec{y'}}
\delta(\vec{y}, \vec{y'})
\theta\left(
\frac{
Q(\vec{y} | \vec{x} )
}{
Q(\vec{y'} | \vec{x'} )
}
< 1
\right )
Q(\vec{x'})
\;.
\label{eq:first-f}
\eeq
Henceforth, we will abbreviate the
probability distributions for the CC's $\classtwo{x}{y}$ and $\classtwo{x'}{y'}$
as follows:

\beq
\pclasstwo{x}{y} \rarrow P_{\rvx, \rvy}
\;\;, \;\;
\pclasstwo{x'}{y'} \rarrow \pwave_{\rvx, \rvy}
\;.
\eeq
Using these abbreviations, one has

\begin{subequations}
\begin{eqnarray}
\theta
\left(
\frac{
Q(\vec{y} | \vec{x} )
}{
Q(\vec{y'} | \vec{x'} )
}
< 1
\right )
&=&
\theta
\left(
\frac{
\exp[ n \sum_{x,y} P(x, y) \ln Q(y |x)]
}{
\exp[ n \sum_{x,y} \pwave(x, y) \ln Q(y |x)]
}
<1
\right)
\\
&=&
\theta
\left(
\sum_{x, y} [P - \pwave](x, y) \ln Q(y |x) <0
\right)
\label{eq:theta-ml-test}
\;.
\eeqa
Substituting Eq.(\ref{eq:theta-ml-test}) into Eq.(\ref{eq:first-f}) and
applying Reduction Formula 2 yields

\begin{eqnarray}
\lefteqn{
f(\vec{x}, \vec{y}) =
\frac{
r(n, N_{\rvx, \rvy} - N_\rvy )
}{
[ \prod\{ P_\rvy \} ]^{\frac{1}{2} (N_\rvx -1)}
}
\int \calD  \pwave_{\rvx, \rvy}\;
\theta(\pwave_{\rvx, \rvy} \geq 0)
\prodset{ \delta( \pwave(y) - P(y) }{y}
}
\nonumber\\
&&
\frac{
\exp[
n H_\pwave(\rvx | \rvy) + n \sum_{x,y} \pwave(x, y) \ln Q(x)
]
}{
\sqrt{ \prodset{ \pwave(x|y)}{x,y}}
}\;
\theta
\left(
\sum_{x, y} [P - \pwave](x, y) L_{xy} <0
\right)
\;.
\end{eqnarray}
Note that

\beqa
\lefteqn{D(\pwave_{\rvx, \rvy}// Q_\rvx \pwave_\rvy )
=
\sum_{x, y}
\pwave(x, y)
\left[
\ln \frac{\pwave(x, y)}{Q(x, y)}
+ \ln \frac{Q(x, y)}{Q(x) Q(y)}
+ \ln \frac{ Q(y)} { \pwave(y) }
\right ]
}
\\
&=&
 D(\pwave_{\rvx, \rvy} // Q_{\rvx, \rvy}) + C_1
+ \sum_{x, y} \Delta \pwave(x, y) L_{xy} - D(\pwave_\rvy //Q_\rvy)
\;.
\eeqa
Hence,

\begin{eqnarray}
f(\vec{x}, \vec{y}) &= &
\frac{
r(n, N_{\rvx, \rvy} - N_\rvy )
\exp[ -n C_1 + n D(P_\rvy // Q_\rvy) ]
}{
[ \prod\{ P_\rvy \} ]^{\frac{1}{2} (N_\rvx -1)}
}
\nonumber\\
&&
\int \calD  \pwave_{\rvx, \rvy}\;
\theta(\pwave_{\rvx, \rvy} \geq 0)
\prodset{ \delta( \pwave(y) - P(y) }{y}
\nonumber\\
&&
\frac{
\exp[
-n D(\pwave_{\rvx, \rvy} // Q_{\rvx, \rvy} )
 - n \sum_{x,y} \Delta \pwave(x, y) L_{xy}
]
}{
\sqrt{ \prodset{ \pwave(x|y)}{x,y}}
}
\nonumber\\
&&
\theta
\left(
\sum_{x, y} [P- \pwave](x, y) L_{xy} <0
\right)
\;.
\label{eq:f-integral-1}
\end{eqnarray}
We will assume that,
 in the integrand of the previous equation,
the inequality constraint  is active; i.e., that
$\sum_{x,y} \Delta P(x, y) L_{xy} =
\sum_{x,y} \Delta \pwave(x, y) L_{xy}$. Therefore, we can
simplify Eq.(\ref{eq:f-integral-1}) by pulling
$e^{-n \sum_{x,y} \Delta \pwave(x, y) L_{xy}}$
outside the integral to get

\begin{eqnarray}\label{eq:f-integral-2}
f(\vec{x}, \vec{y}) &= &
\frac{
r(n, N_{\rvx, \rvy} - N_\rvy )
\exp[ -n C_1 + n D(P_\rvy // Q_\rvy)
- n \sum_{x,y} \Delta P(x, y) L_{xy}
]
}{
[ \prod\{ P_\rvy \} ]^{\frac{1}{2} (N_\rvx -1)}
}
\nonumber\\
&&
\int \calD  \pwave_{\rvx, \rvy}\;
\theta(\pwave_{\rvx, \rvy} \geq 0)
\prodset{ \delta( \pwave(y) - P(y) }{y}
\nonumber\\
&&
\frac{
\exp[
-n D(\pwave_{\rvx, \rvy} // Q_{\rvx, \rvy} ) ]
}{
\sqrt{ \prodset{ \pwave(x|y)}{x,y}}
}
\nonumber\\
&&
\theta
\left(
\sum_{x, y} [P- \pwave](x, y) L_{xy} <0
\right)
\;.
\end{eqnarray}
To find $\ln(f)$ to leading order in $n$, we need to find
the  point $\pwave^*(x,y)$ that dominates the
integral on the RHS of Eq.(\ref{eq:f-integral-2}). To find $\pwave^*$, we must
minimize the following Lagrangian with respect to
$\pwave, \lam$, and $ \mu_y$:

\beq
\calL = D(\pwave_{\rvx, \rvy}// Q_{\rvx, \rvy})
-\lam \left ( \sum_{x, y} (P - \pwave)(x, y) L_{xy} \right)
+\sum_y \mu_y ( P - \pwave)(y)
\;.
\label{eq:lagr-noisy-exact}
\eeq
The Gaussian approximation for the previous Lagrangian is:

\beq
\calL =
\sum_{x, y} \frac{[ \Delta \pwave(x, y)]^2}{2 Q(x, y)}
-\lam \left ( \sum_{x, y} (P - \pwave)(x, y) L_{xy} \right)
+\sum_y \mu_y ( P - \pwave)(y)
\;.
\label{eq:lagr-noisy-quad}
\eeq
Assume that the exact Lagrangian of Eq.(\ref{eq:lagr-noisy-exact}) is
well approximated
by its Gaussian approximation. (This assumption is not
necessary and will be removed later, in Appendix \ref{app:t-exp}.)
Let

\beq
\alpha_{xy} = L_{xy} - \sum_{x'} Q(x'|y) L_{x'y}
\;,
\eeq

\beq
\av{\alpha} = \sum_{x,y} Q(x,y) \alpha_{xy} = 0
\;,
\eeq

\beqa
\av{\alpha^2 } &=& \sum_{x, y} Q(x, y) \alpha_{xy}^2\\
&=& \sum_{x, y} Q(x, y) L_{xy} \alpha_{xy}
\;.
\eeqa
Minimizing Eq.(\ref{eq:lagr-noisy-quad}) with respect to
$\pwave, \lam$, and $\mu_y$ yields

\beq
\lam = \frac{ - \sum_{x, y} P(y)\Delta P(x| y) L_{xy}
}{
\av{\alpha^2 }}
\;,
\eeq
and

\beq
\Delta \pwave^*(x, y) = -\lam Q(x, y) \alpha_{xy} - Q(x|y) \Delta P(y)
\;.
\eeq
If $\calL^*$ is the value of $\calL$ at the extremum, then

\beq
\calL^* = \sum_y
\frac{
[\Delta P(y)]^2
}{
2 Q(y)
}
+t
\;,
\eeq
where, to lowest order in $\Delta P$, $t$ is given by

\beq
t =
\frac{
\epsilon^2
}{
2 \av{\alpha^2}
}
\;,
\label{eq:leading-t}
\eeq
where

\beq
\epsilon = \sum_{x, y} P(y)\Delta P(x| y) L_{xy}
\;.
\eeq

Now that we know $\calL^*$, we can apply Laplace's Method
to the integral on the RHS of Eq.(\ref{eq:f-integral-2}) to get

\beqa
\ln(f) &\approx &
-nC_1 + n D(P_\rvy // Q_\rvy) - n \sum_{x, y} \Delta P(x, y) L_{xy}  -n \calL^*\\
&\approx & -n\left(C_1 + \sum_{x, y}\Delta P(x, y) L_{xy} + t\right)
\;.
\eeqa
This value for $\ln(f)$ can be inserted into Eqs.(\ref{eq:perr-v-M})
and (\ref{eq:v-M-theta}) to get

\begin{eqnarray}
p_{err}
&\approx&
1 - r(n, N_{\rvx, \rvy})
\int_{pd(S_{\rvx,\rvy})} \calD P_{\rvx, \rvy}\;
\nonumber \\
&&
\frac{\exp[-n D(P_{\rvx,\rvy} // Q_{\rvx, \rvy})]}
{
\sqrt{\prod \{ P_{\rvx,\rvy} \}}
}
\theta(\Delta R - \sum_{x,y} \Delta P(x, y) L_{xy} - t<0)
\;.
\label{eq:perr-noisy-with-t}
\end{eqnarray}

Assume that the integral of the previous equation has been rescaled so
that its integrand is in the Gaussian regime. Then

\begin{eqnarray}\label{eq:perr-noisy-without-t}
p_{err}
&\approx&
1 -
\frac{r(n, N_{\rvx, \rvy})
}{
\sqrt{\prod \{ P^*_{\rvx,\rvy} \}}
}
\int_{pd(S_{\rvx,\rvy})} \calD P_{\rvx, \rvy}\;
\nonumber \\
&&
\exp[-n \sum_{x,y} \frac{[\Delta P(x, y)]^2}{2 Q(x,y)}]
\theta(\sum_{x,y} \Delta P(x, y) L_{xy} \geq \Delta R )
\;.
\end{eqnarray}
Let

\beq
\beta_{xy} =L_{xy} - \sum_{x,y} Q(x,y) L_{xy}=L_{xy} - C_1
\;,
\eeq

\beq
\av{\beta} = \sum_{x, y} Q(x, y)  \beta_{xy}=0
\;,
\eeq

\beq
\av{\beta^2} = \sum_{x, y} Q(x, y)  \beta_{xy}^2
\;.
\eeq
Applying the Gaussian Integration Formulae of Appendix
\ref{app:gaussian-int} to the RHS of Eq.(\ref{eq:perr-noisy-without-t})
yields

\beq
p_{err} \approx 1-
\frac{1}{2 u} \erfc
\left(
\Delta R
\sqrt{\frac{n}{2 \av{\beta^2}}}
\right)
\;,
\label{eq:perr-erfc-noisy}
\eeq
where

\beq
u = \sqrt{
\frac{
\prod\{P^*_{\rvx, \rvy} \}
}{
\prod\{ Q_{\rvx, \rvy} \}
}
}
\;.
\label{eq:u-noisy}
\eeq
To find the dominant point $P^*_{\rvx, \rvy}$
alluded to in Eq.(\ref{eq:u-noisy}),
one must minimize the following Lagrangian
with respect to $P, \lam$ and $\mu$:

\beq
\calL = \sum_{x,y} \frac{ [\Delta P(x,y)]^2 }{ 2Q(x,y) } +
\lam ( \Delta R - \sum_{x,y} \Delta P(x,y) L_{xy} ) + \mu
\sum_{x,y} \Delta P(x,y) \;.
\eeq
One finds that the extremum is
at

\beq
\Delta P^*(x,y)=
B(x,y)\Delta R
\;,
\eeq
where

\beq
B(x,y) = \frac{\beta_{xy}Q(x, y)}{\av{\beta^2}}
\;.
\eeq
Substituting this value for $\Delta P^*$
into Eq.(\ref{eq:u-noisy})   gives

\beq
u = \sqrt{
\prodset{ 1 +
\frac{ \beta_{xy} \Delta R }
{ \av{\beta^2} }
}{(x,y)}
}
\;.
\eeq

Note that this paper has exposed
a close analogy between noiseless and
noisy coding, as far as $p_{err}$ is concerned.
For example,
Eq.(\ref{eq:perr-nless}) for noiseless coding
is analogous to
Eq.(\ref{eq:perr-noisy-with-t}) for noisy coding.
Likewise,
Eq.(\ref{eq:perr-erfc-nless}) is analogous to Eq.(\ref{eq:perr-erfc-noisy}).

\subsection{RG Equations}\label{sec:new-noisy-rg-eqs}
In this section, we will calculate
the RG equations for channel transmission
using random encoding and ML decoding.

For noiseless coding, the RG equations arose
from rescaling Eq.(\ref{eq:perr-nless}).
In the case we are now considering, that of noisy coding,
the RG equations arise from rescaling
Eq.(\ref{eq:perr-noisy-with-t}). Note the close resemblance
between these two equations.

In the noiseless coding case, we found
a RG equation for
$P_\rvx$  by assuming that the argument $nD(P_\rvx//Q_\rvx)$
of the exponential in the integrand of Eq.(\ref{eq:perr-nless})
was invariant under a change of scale.
In analogy, for noisy coding, we find a RG for
$P_{\rvx, \rvy}$ by assuming that
the argument $nD(P_{\rvx, \rvy}//Q_{\rvx, \rvy})$
of the exponential in the integrand
of Eq.(\ref{eq:perr-noisy-with-t}) is invariant under
a change of scale.  We get

\beq
\pder{ \Delta P^\sfun(x)}{s}
=
-\gamma_0(P^\sfun, Q)
\Delta P^\sfun(x)
\;,
\eeq
where

\beq
\gamma_0(P, Q) =
\frac{
D(P//Q) }{
D(P//Q) + D(Q//P)
}
\;.
\eeq

In the noiseless coding case, we found
a RG equation for
$\Delta R$  by assuming that
the theta function in the integrand of Eq.(\ref{eq:perr-nless})
was invariant under a change of scale.
In analogy, for noisy coding, we find a RG for
$\Delta R$ by assuming that
the theta function in the integrand
of Eq.(\ref{eq:perr-noisy-with-t}) is invariant under
a change of scale.  We get

\beq
\pder{ \Delta R^\sfun}{s} = -\gamma_1( P^\sfun, Q) \Delta R^\sfun
\;,
\eeq
where

\beq
\gamma_1(P, Q) =
\lim_{s\rarrow 0}
\frac{
(-1) \pder{T(P^\sfun, Q)}{s}
}{
T(P^\sfun, Q)
}
\;,
\label{eq:gamma1-noisy}
\eeq
where

\beq
T(P, Q) = T_0 + t
\;,
\label{eq:T-noisy}
\eeq
where

\beq
T_0= \sum_{x,y} \Delta P(x,y) L_{xy}
\;.
\eeq
For any real valued function $f(s)$ of $s\geq 0$, define

\beq
Df = \lim_{s\rarrow 0} \left(\frac{-1}{\gamma_0}\right)\pder{f}{s}
\;.
\eeq
Note that $DP^\sfun = \Delta P$ and $\gamma_1 = \gamma_0 \frac{DT}{T}$.
Substituting Eq.(\ref{eq:T-noisy})
into Eq.(\ref{eq:gamma1-noisy}) gives

\beq
\gamma_1(P, Q) =
\left(
1 +
\frac{-t + Dt}{ T}
\right)
\gamma_0(P, Q)
\;.
\label{eq:gamma-one-from-t-Dt}
\eeq
Eq.(\ref{eq:leading-t}) gives $t$ to lowest order
in $\epsilon$. It is easy to show that for such a $t$,
$Dt = 2t$, so $\gamma_1 = ( 1 + \frac{t}{T})\gamma_0$.
In Appendix \ref{app:t-exp}, we find  $t$ and $\gamma_1$
to all orders in $\epsilon$.

\subsection{Coda to Error Model}

It is customary \cite{cover-thomas}
to end a discussion of noisy coding with random encoding with the
following 3 observations.

\begin{description}
\item[Replace $C_1$ by Capacity.]
In $C_1$, $Q(x)$ and $Q(y|x)$ are independent.
The capacity is defined by $C= \max_{Q_\rvx\in pd(S_\rvx)} C_1$.
Let  $Q^*_\rvx\in pd(S_\rvx)$ be the probability distribution $Q_\rvx$
that maximizes $C_1$ at fixed $Q(y|x)$.
The $p_{err}$ that we derived for random encoding depends on $C_1$.
It is advantageous to set $Q_\rvx = Q^*_\rvx$ in $p_{err}$
since $p_{err}(C) \leq p_{err}(C_1)$.

\item[Keep Best Codebook.]
The  $p_{err}$
that we derived for random encoding
was averaged over all
possible  codebooks $\kappa$ (there are $N_\rvx^{nM}$ of them).
There must exist a ``best" codebook $\kappa_{best}$ among these such that
$p_{err}(\kappa_{best})\leq p_{err}(\kappa)$ for all $\kappa$,
and therefore $p_{err}(\kappa_{best})\leq$  mean of $(p_{err}(\kappa))_\kappa$.

\item[Keep Ruly Half of Codebook.]

Suppose $x_1\leq x_2\leq \ldots \leq x_N$ is a
monotonically non-decreasing sequence of real numbers.
Define partial sums $S_{a,b} = x_a + x_{a+1} + \ldots + x_b$ for $a\leq b$.
The mean of the sequence is $\mu = S_{1, N}/N$ and
its median is $x_{\frac{N}{2}}$. It is easy to prove by contradiction
that
$x_{\frac{N}{2}}\leq 2\mu$.

Define  the ``unruly half" $S_\rvm^{unruly}$ of a codebook to be
the set of all $m\in S_\rvm$ for which $p_{err|m}$ is larger than the
median of $(p_{err|m})_{\forall m\in S_\rvm}$. Thus,
$S^{ruly}_\rvm \cup S^{unruly}_\rvm = S^{all}_\rvm$. If we remove
the ``unruly half "of a codebook, then we end up with a
new codebook with half as big an $M$; symbolically, $M_{ruly} =
\frac{M_{all}}{2}$. In the limit of large codeword size $n$, this
does not affect the rate $R$ too much. Indeed, $R_{ruly} =
\frac{1}{n} \ln ( \frac{M_{all}}{2}) = R_{all} - \frac{1}{n}
\ln(2) \rarrow R_{all}$. The advantage of keeping only the ruly
half of a codebook is that $p_{err|m}$ for all $m\in
S_\rvm^{ruly}$ is bounded above by $2 p_{err}(all)$.

\end{description}

\section{Computer Results}\label{sec:comp-results}

In this section, we will describe the algorithms used by
the computer program
WimpyRG-C1.0 to solve the equations of this paper,
and we will give
examples of typical inputs and outputs of said program.
For more information about WimpyRG, see its source code and
accompanying documentation.

\subsection{Old-Noiseless  Approximation of $p_{err}$}
First, let us describe how WimpyRG
calculates the old fashioned approximation for $p_{err}$,
in the case of noiseless coding.

We shall indicate derivatives by primes.
Previously, we defined

\beq
Z(\lam) = \sum_x Q(x)^{\frac{1}{1+\lam}}
\;,
\eeq

\beq
\gamma(\lam)  = \lam R - (1+\lam) \ln Z(\lam)
\;,
\eeq

\beq
\gamma = \max_{\lam\geq 0} \gamma(\lam)
\;,
\eeq
and
we showed that the
probability of error is approximated by

\beq
p_{err} = e^{-n \gamma}
\;.
\label{eq:wimpy-old-nless}
\eeq

To maximize the function $\gamma(\lam)$,
WimpyRG uses the simple
Newton Raphson (NR) method as follows.
Note that only the range
$R\in (0, \ln N_\rvx)$ is of interest.
It is easy to show that for all $\lam\geq 0$,
if $R\in (H(Q) , \ln N_\rvx)$, then $\gamma(\lam)$ has a negative second derivative and
$\primeone{\gamma}(0) = \Delta R> 0$. Hence,
for $R\in (H(Q), \ln N_\rvx)$, $\gamma(\lam)$ has a
unique maximum at some point $\lam = \lam_0>0$.
The NR method is way of
finding the zeros of a function $f: Reals\rarrow Reals$. Suppose
that $f(x)=0$ at $x=a$. We can Taylor expand $f(x)$
to first order about this zero:
$f(x) \approx  f(a) + f^{\; \prime}(a) (x-a)$.
Thus,
$f(x)=0$ implies $x = a -f(a)/\primeone{f}(a)$.
This suggest the recursion relation:
$x_{n+1} = x_n - f(x_n)/\primeone{f}(x_n)$ for $n=0, 1, 2, \ldots$.
Replacing  $x$ by $\lam$, and $f(x)$ by $\primeone{\gamma}(\lam)$,
one gets

\beq
\lam_{n+1} = \lam_n - \frac{\primeone{\gamma}(\lam_n)}{\primetwo{\gamma}(\lam_n)}
\;.
\eeq
WimpyRG uses the previous recursion relation
to find the maximum of $\gamma(\lam)$.
This algorithm requires that we know
the functions $\primeone{\gamma}(\lam)$ and  $\primetwo{\gamma}(\lam)$.
These two derivatives can be computed explicitly as follows.
Define

\beq
Z_n(\lam) = \sum_x Q(x)^{\frac{1}{1+\lam}} [\ln Q(x) ]^n
\;.
\eeq
Note that $Z = Z_0$.
It is easy to show that

\beq
\primeone{\gamma}(\lam) =
R - \ln Z_0 + \frac{Z_1}{(1+\lam) Z_0}
\;,
\eeq
and

\beq
\primetwo{\gamma}(\lam) = \frac{-(Z_0 Z_2 - Z_1^2)}{ (1+\lam)^3 Z^2_0}
\;.
\eeq

\subsection{New-Noiseless and New-Noisy  Approximations of $p_{err}$}
Next, let us describe how WimpyRG
calculates the new (CCRG) approximation for $p_{err}$,
in the case of either noiseless or noisy coding.

For both noiseless and noisy coding, we must solve the following pair of
coupled RG equations.
For $s\geq 0$,

\beq
\pder{ \Delta R^\sfun}{s} = -\gamma_1( P^\sfun, Q) \Delta R^\sfun
\;,
\label{eq:wimpy-delr-diff-eq}
\eeq
and

\beq
\pder{ \Delta P^\sfun(X)}{s}
=
-\gamma_0(P^\sfun, Q)
\Delta P^\sfun(X)
\;
\label{eq:wimpy-delp-diff-eq}
\eeq
for all $X\in S_\rvX$, where $S_X = S_\rvx$ for noiseless coding and
$S_\rvX = S_{\rvx, \rvy}$ for  noisy coding.
We must solve this pair of RG equations subject to the
following pair of boundary conditions:
At $s=0$:
\beq
\Delta R^{(0)} = \Delta R
\;,
\label{eq:wimpy-delr-bc}
\eeq
and
at  $s= \sfin$:

\beq
P^{ (\sfin)} (X) = Q(X) + B(X)\Delta R^{(\sfin)}
\;,
\label{eq:wimpy-delp-bc}
\eeq
for all $X$.
$\gamma_0$ and $\gamma_1$ are known functions of $P$ and $Q$.
$\gamma_0$ is the same for both noiseless and noisy coding, but
$\gamma_1$ is different.
$\Delta R$ is assumed to be known. $\Delta R$ equals $R-H(Q)$ for noiseless coding
and $R - C_1$ for noisy coding.
The test fractions $\Phi_0(P, Q)$ and $\Phi_1(P, Q)$
are also known functions of $P$ and $Q$.
 $\sfin$ is defined as any $s$ large enough for the following
to be true: $\Phi_0(P^{(\sfin)}, Q)<<1$ and $\Phi_1(P^{(\sfin)}, Q)<<1$.
$B(x)$ is also a known function. It depends on $Q$ but not $P$, and
it differs for noiseless and noisy coding.

Eqs.(\ref{eq:wimpy-delr-diff-eq})
and (\ref{eq:wimpy-delp-diff-eq}) can be solved recursively
by performing the following steps:

\begin{description}

\item[(1) Move Backwards (from $s= \sfin$ to $s=0$)]
This step will be performed either at the beginning of the
algorithm, or after performing step (2) below.
If this step is being performed after step (2),
then step (2) has just yielded a fresh value
of $\Delta R^\sfinfun$.
On the other hand, if this step is being performed at the beginning of
the algorithm, take $\Delta R^\sfinfun = 10^{-12}$. \cite{first-Del-R}

Substituting $\Delta R^{(\sfin)}$ into
Eq.(\ref{eq:wimpy-delp-bc}) gives $\Delta P^\sfinfun$. Hence we can
solve Eq.(\ref{eq:wimpy-delp-diff-eq}) numerically (using
the Fourth Order Runge Kutta algorithm  \cite{AbraSteg} ) to get $\Delta P^\sfun (x)$
for all $x\in S_\rvx$ and all $s\in [0, \sfin]$. These $\Delta
P^\sfun(x)$ values can in turn be used to calculate
 $\gamma_1(P^\sfun, Q)$ for each $s\in[0,\sfin]$.

\item[(2) Move Forwards (from $s=0$ to $s= \sfin$)]
After step (1), we have a fresh value of
$\gamma_1(P^\sfun, Q)$ for each $s\in[0,\sfin]$.
By virtue of Eq.(\ref{eq:wimpy-delr-bc}), $\Delta R^{(0)}$
is also known. Hence we can solve Eq.(\ref{eq:wimpy-delr-diff-eq}) numerically
(again,
using the Fourth Order Runge Kutta algorithm ) to get
$\Delta R^{(\sfin)}$.

\end{description}

One performs steps (1), (2), (1), (2), ...., until the
difference between two successive values of
$\Delta R^{(\sfin)}$ is very small.

Let

\beq
\calE = \frac{1}{2}\erfc \left( \Delta R^\sfinfun
\sqrt{\frac{n^\sfinfun}{2 \av{\beta^2}}} \right)
 \;,
\label{eq:wimpy-E} \eeq
 where $n^\sfinfun = e^\sfin n$. The
probability of error $p_{err}$ is approximately equal to $\calE $ for
noiseless coding and to $1-\calE $ for noisy coding. However, the
quantities $\Delta R^\sfinfun$ and $ \av{\beta^2}$ that appear in
$\calE $ have different definitions for noiseless and noisy coding.

\subsection{Examples of WimpyRG Input and Output}

\begin{figure}[h]
    \begin{center}
  \epsfig{file=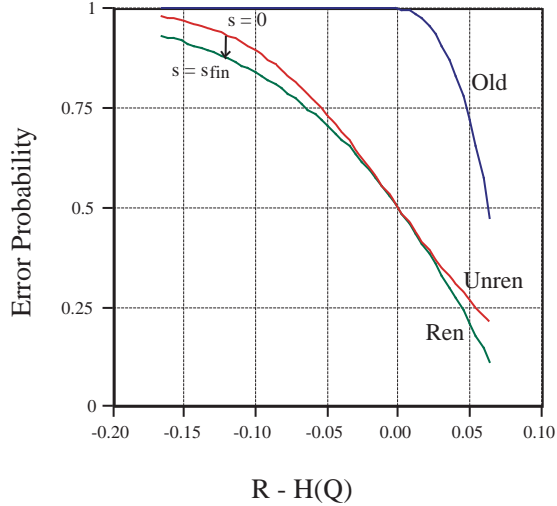, height=3.0in}
    \caption{A plot of WimpyRG output for noiseless coding.}
    \label{fig:R-Prob-graph-nless}
    \end{center}
\end{figure}

Fig.\ref{fig:R-Prob-graph-nless} is a plot of WimpyRG output
for noiseless coding. It gives
$p_{err}$ as a function of $R- H(Q)$, for
 $n=20$
and
$\vec{Q}_\rvx= (
.20,
.30,
.13,
.37)$.
$H(Q)= 1.316$. The maximum possible $R$ is $\ln(N_\rvx) = 1.386$.
Curve \old, the old approximation of $p_{err}$,
is a plot of Eq.(\ref{eq:wimpy-old-nless}).
Let $\calE $ be given by Eq.(\ref{eq:wimpy-E}).
Curve \unren, the unrenormalized approximation of  $p_{err}$,
is a plot of $\calE $
 with $\sfin=0$ (hence $n^\sfinfun = e^\sfin n = 20$).
Curve \ren, the renormalized approximation of  $p_{err}$,
is a plot of $\calE $
with $\sfin=7.5$ (hence $n^\sfinfun = e^\sfin n = 36160.8
$.)

It appears from Fig.\ref{fig:R-Prob-graph-nless} that
curve \unren  is always higher or equal
to curve \ren . As expected, both the \old and \ren curves plummet
towards $p_{err}=0$ at $R = \ln N_\rvx$.

 Curve \old  is not expected to be a good
approximation for $p_{err}$ when $R$ is close to $H(Q)$.
Indeed, for  $R=H(Q)$, $\gamma = 0$, so $e^{-n \gamma}$
is indeterminate
because $n\gamma =\infty \cdot 0$ .  On the other hand,
curve \ren
 is expected to behave best when $R$ is
near $H(Q)$, in the sense that the closer $R$ is to $H(Q)$, the
lower the value of $\sfin$ that is required to reach the
quadratic regime.

While generating the points  $(\Delta R, p_{err})$
plotted in Fig.\ref{fig:R-Prob-graph-nless},
WimpyRG also generated certain
figures of merit for each point. For
example,
when generating the point
$(\Delta R, p_{err})=(-0.15825, 0.925769)$,
WimpyRG
also generated:

\begin{verbatim}
====================
number of cycles (max is 100) = 6
test fraction 0 (initial, final) = 0.15137, 0.00234084
test fraction 1 (initial, final) = 0.397863, 0.0105788
n (initial, final) = 20, 36160.8
Delta R (initial, final) = -0.15825, -0.00271302
R, unrenormalized error_prob, error_prob = 1.15793, 0.976272, 0.925769
====================
\end{verbatim}
In this output, ``initial" always refers to $s=0$ and ``final" to
$s=\sfin= 7.5$. A ``cycle" is defined as a single application
of the Backward/Forward steps defined previously.
A cycle takes
the computer program from $s=\sfin$ to $s=0$ and back again.
The ``number of cycles" is how many cycles were required
before reaching a  reasonably constant  (i.e. varying no more than 0.1\%
between successive cycles)
value
 for $\Delta R^\sfinfun$.
Notice that test fractions $\Phi_0$ and $\Phi_1$
decreased
substantially whereas $n$ increased substantially in going from $s=0$ to $s=\sfin$.
Hurray!

\begin{figure}[h]
    \begin{center}
  \epsfig{file=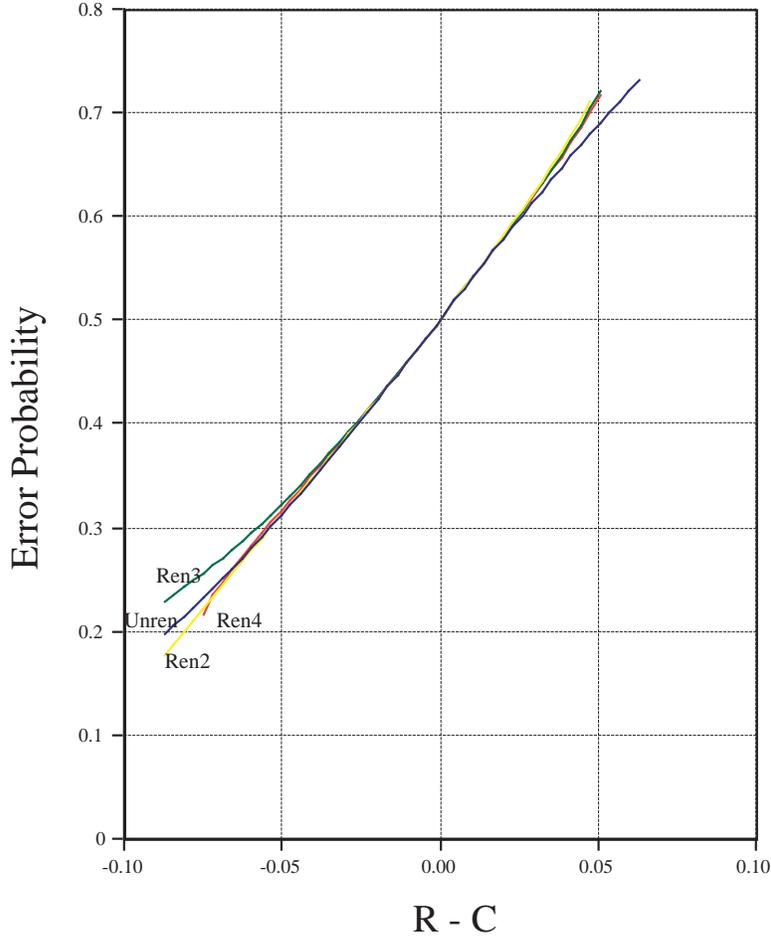, height=5.0in}
    \caption{A plot of WimpyRG output for noisy coding.}
    \label{fig:R-Prob-graph-noisy}
    \end{center}
\end{figure}

\begin{figure}[h]
    \begin{center}
   \epsfig{file=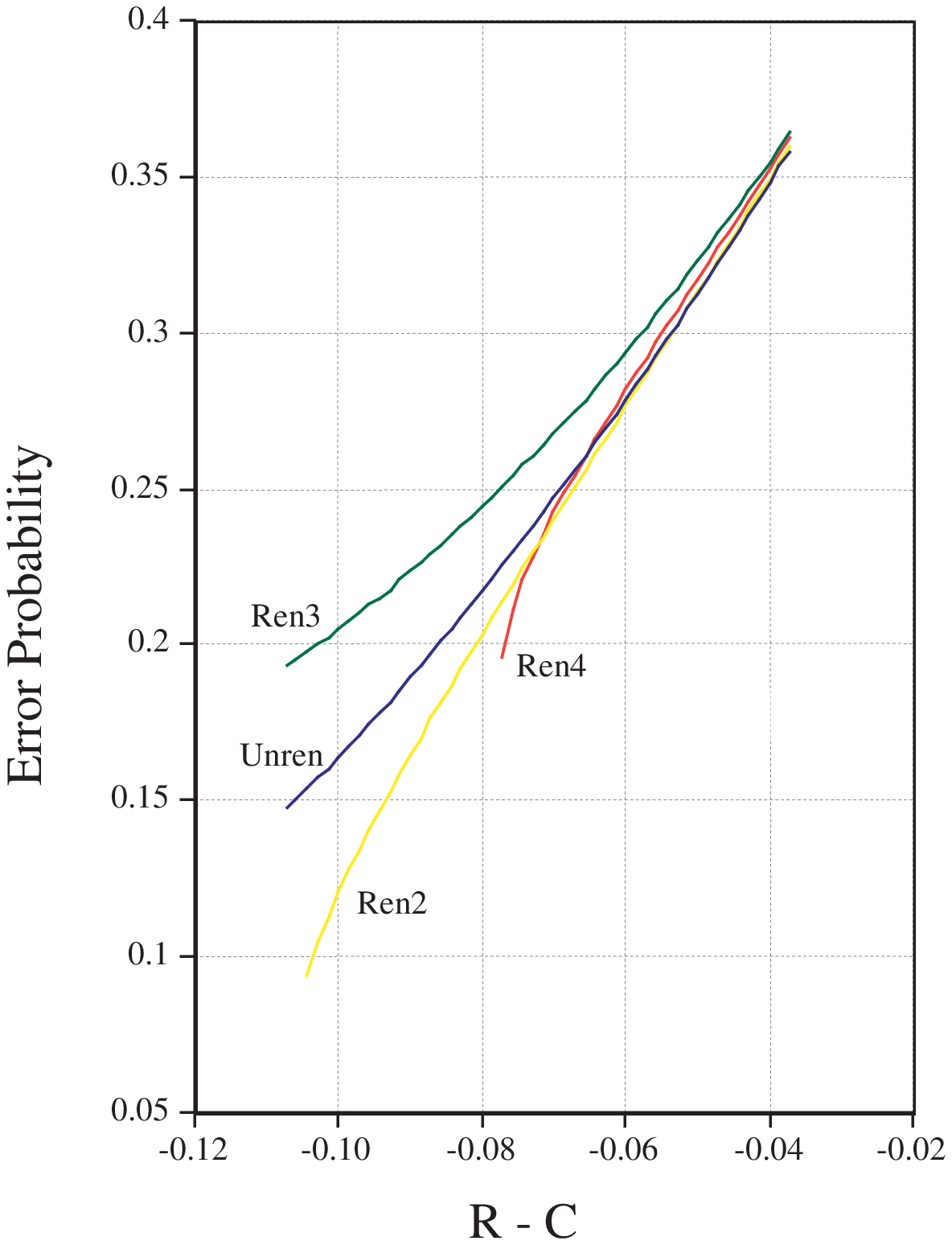, height=4.0in}
    \caption{Magnified view of part of Fig.\ref{fig:R-Prob-graph-noisy},
the part with the smallest $\Delta R$ values.}
    \label{fig:mag-view-R-Prob-graph-noisy}
    \end{center}
\end{figure}

Fig.\ref{fig:R-Prob-graph-noisy} is a
plot of WimpyRG output for noisy coding.
It gives  $p_{err}$ as a function of $R- C$, for
 $n=20$. The channel probability $Q(y|x)$ for these plots
is $Q(0|0)= Q(1|1) = 0.99$, $Q(1|0) = Q(0|1) = 0.01$ (a
symmetric binary channel). The source distribution $Q(x)$ is
$Q(0) = Q(1)= 0.5$, as required to make $C_1=C$
for a binary symmetric channel. For this $Q(y|x)$
and $Q(x)$, $C= 0.637 nats$ (or $C=.919 bits$ if one uses base 2
logs). Let $\calE $ be given by Eq.(\ref{eq:wimpy-E}). Curve \unren, the
unrenormalized approximation of  $p_{err}$, is a
plot of $1-\calE $ with $\sfin=0$ (hence $n^\sfinfun = e^\sfin n =
20$). Curves \rendos, \rentres and \rencua,
renormalized approximations of
$p_{err}$, are plots of $1-\calE $ with $\sfin=7.5$ (hence $n^\sfinfun
= e^\sfin n = 36160.8 $.)
To obtain curve \ren j for $j\in\{2,3,4\}$,
we used an approximation for $t$
that included terms up to and including order $\epsilon^j$.
See Appendix \ref{app:t-exp}.

Fig.\ref{fig:mag-view-R-Prob-graph-noisy} is
a magnified view of a part of
Fig.\ref{fig:R-Prob-graph-noisy}, the part with the smallest values of
$\Delta R$.
Each renormalized curve \ren j for $j\in\{2,3,4\}$
has endpoints $a_j$ and $b_j$ such that the curve is shown
only for $\Delta R \in [a_j,b_j]$.
We found that our  algorithm for obtaining \ren j
broke down for $\Delta R < a_j$ and $\Delta R>b_j$.
There is no guarantee that the Runge Kutta algorithm
that we use for solving the RG equations will not produce
unphysical values such as a $P^\sfun(X)\not\in [0,1]$
or a $\gamma_1 <0$ at some intermediate step.
Such unphysical  values for $P^\sfun(X)$ or $\gamma_1$
were obtained by WimpyRG for $\Delta R<a_j$ or
$\Delta R>b_j$ but not for $a_j < \Delta R < b_j$.
We conjecture that a curve \ren$\infty$ that
used $t$ to all orders in $\epsilon$ would
 reach $p_{err}=0$
and $p_{err} =1$ at finite values of $\Delta R$.

\newpage
\appendix
\section{Appendix: Error Function}\label{app:err-fun}
This appendix reviews well known properties
of the Error Function\cite{AbraSteg}.

\begin{figure}[h]
    \begin{center}
    \epsfig{file=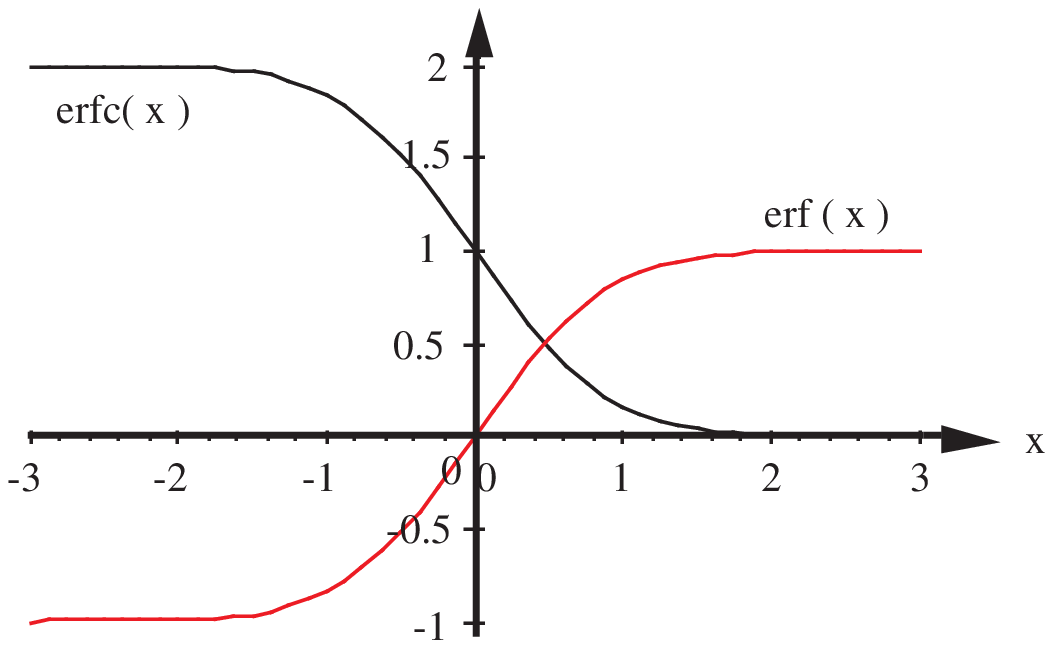, height=2in}
    \caption{Plot of erf($x$) and erfc($x$).}
    \label{fig:erf-plot}
    \end{center}
\end{figure}

The Error Function is defined for real $x$ by

\beq
\erf(x) = \frac{2}{\sqrt{\pi}} \int_0^x d\xi \; e^{-\xi^2}
\;.
\eeq
$\erf(x)$ can be analytically continued to complex $x$,
but we have no need to consider such an extension in this paper.
The complement of the Error Function is defined by

\beq
\erfc(x) = 1 - \erf(x) = \frac{2}{\sqrt{\pi}} \int_x^\infty d\xi \; e^{-\xi^2}
\;.
\eeq
See  Fig.\ref{fig:erf-plot} for a plot of erf($x$) and erfc($x$).
Under reflection $x\rarrow -x$, erf($x$) obeys

\beq
\erf(-x) = - \erf(x)
\;,
\eeq
and erfc() obeys

\beq
\erfc(-x) = 1 - \erf(-x) = 2 - \erfc(x)
\;.
\eeq
For real $x$ such that $|x|<<1$,

\beq
\erf(x) = \frac{2}{\sqrt{\pi}} \left( x
- \frac{x^3}{3\cdot 1!}
+ \frac{x^5}{5\cdot 2!}
- \frac{x^7}{7\cdot 3!}
+ \ldots
\right)
\;.
\eeq
For real $x$ such that $|x|>>1$,

\beq
\erfc(x) = 2 \theta(x<0) +
\frac{e^{-x^2}}{x \sqrt{\pi}}\left(
1 - \frac{1}{2x^2}
+ \frac{1\cdot 3}{(2x^2)^2}
- \frac{1\cdot 3\cdot 5}{(2x^2)^3}
+ \ldots \right)
\;.
\eeq

\begin{claim}
For $a, b, \Lam \in Reals$ with $\Lam, a>0$,

\beq
\erfc( \frac{b}{2 \sqrt{a}}) = \frac{1}{\pi i}
\int_{\Lam - i\infty}^{\Lam + i \infty}
\frac{d\lam}{\lam}\; \exp(a \lam^2 - b \lam)
\;.
\label{eq:erfc-contour-int-a-b}
\eeq
\end{claim}
{\bf proof:}

\begin{subequations}
\label{eq:erfc-contour-int}
\begin{eqnarray}
\erfc(x)
& = &
 \frac{2}{\sqrt{\pi}}
 \int_{-\infty}^{+\infty} d\xi \; e^{-\xi^2} \theta(\xi > x)\\
&=&
 \frac{1}{\pi^{\frac{3}{2}} i} \int_{\Lam - i\infty}^{\Lam + i\infty}
\frac{d\lam}{\lam}
\int_{-\infty}^{+\infty} d\xi \; \exp(-\xi^2 + \lam\xi  - \lam x)\\
&=&
\frac{1}{\pi i}
\int_{\Lam - i\infty}^{\Lam + i\infty}
\frac{d\lam}{\lam} \;
 \exp(\frac{\lam^2}{4} -\lam x)
\;.\label{eq:erfc-contour-int-c}
\eeqa
In Eq.(\ref{eq:erfc-contour-int}), we went
from line (a) to (b) by using the integral representation
of the theta function, as given by Eq.(\ref{eq:theta-int-rep}).
Now make the replacements
$\lam \rarrow 2 \sqrt{a} \lam$,
$x\rarrow \frac{b}{2 \sqrt{a}}$
in Eq.(\ref{eq:erfc-contour-int-c}).
QED

\section{Appendix: Taylor Expansions Related to Information Theory  }
\label{app:taylor-exp} This handy appendix collects in one place
several Taylor expansions that arise frequently in Information
Theory.

For real $x$ such that $|x|<1$,

\beqa
\ln (1 + x)
&=&
\sum_{n=1}^\infty \frac{ (-1)^{n+1}x^n}{n}\\
&=&
x - \frac{x^2}{2} + \frac{x^3}{3} + \ldots
\;.
\eeqa
Thus, for $|\frac{h}{x}|<1$,

\beqa
\ln(x + h)
&=& \ln[ x( 1 + \frac{h}{x})]
= \ln x  + \sum_{n=1}^\infty \frac{ (-1)^{n+1}(\frac{h}{x})^n}{n}
\\
&=& \ln x + \frac{h}{x} - \frac{h^2}{2x^2} + \ldots
\;,
\eeqa

\beqa
(x + h) \ln (x +h)
&=&
x\ln x + h( \ln x + 1) + h \sum_{n=2}^\infty \frac{(-1)^n }{n(n-1)}\left( \frac{h}{x}\right)^{n-1}\\
&=&
x\ln x + h( \ln x + 1)  + \frac{h^2}{2 x} + \ldots
\;.
\eeqa
Let
$\Delta P(x) = P(x) - Q(x)$. Then
\beqa
H(P) &=&
-\sum_x P(x) \ln P(x) \\
&=&
H(Q) - \sum_x \Delta P(x) \ln Q(x)
-
\sum_x \frac{[\Delta P(x)]^2}{2 Q(x)} + {\cal O}(( \Delta P )^3)
\;,
\eeqa
and

\beqa
D(P//Q) &=&
 \sum_x P(x) \ln \frac{P(x)}{Q(x)}\\
& = &
\sum_x \frac{[\Delta P(x)]^2}{2 Q(x)} + {\cal O}(( \Delta P )^3)
\;.
\eeqa

\section{Appendix: Gaussian Integration Formulae}\label{app:gaussian-int}
In this  appendix, we present certain integration formulae that
contain a Gaussian times a delta or a theta function in the integrand.

The following lemma will be used to prove Claim \ref{th:gau-ints},
which is the main result of this appendix.

\begin{lemma}\label{lem:inv-det-B}
Suppose $A\in Reals^{n\times n}$ is invertible,
$v\in Reals^{n\times 1}$, $v^T A^{-1} v \neq 0$,
$0<\epsilon<<1$, and

\beq
B = A + \frac{vv^T}{\epsilon}
\;.
\eeq
Then the inverse and determinant of B are given by

\beq
B^{-1} =
A^{-1}
-
A^{-1} \tilde{A} A^{-1}
\;\;{\rm where} \;\;
\tilde{A} =
\frac{vv^T}{v^T A^{-1}v}
\;,
\label{eq:inv-B}
\eeq
and

\beq
\det B =
\det(A)
\frac{v^T A^{-1} v}{\epsilon}
\;.
\label{eq:det-B}
\eeq

\end{lemma}
{\bf proof:}

It is easy to show that if $u$ and $v$ are $n$ dimensional column vectors
and

\beq
B = A +  uv^T
\;,
\eeq
then
\beq
B^{-1} =
A^{-1} - \frac{ A^{-1} uv^T A^{-1}}{ 1 + v^T A^{-1} u}
\;
\eeq
satisfies $B B^{-1} = B^{-1} B = 1$.
Setting $u = v/\epsilon$ and taking the limit
$\epsilon\rarrow 0$ yields Eq.(\ref{eq:inv-B}).

To show Eq.(\ref{eq:det-B}), recall that

\beq
\ln (\det A) = \tr (\ln A)
\;.
\label{eq:ln-det-eq-tr-ln}
\eeq
(This well known identity is obvious when $A$ is diagonal.
The proof is also very simple when $A$ is non-diagonal but
diagonalizable.) If the entries of $A$ are taken to be independent
variables, then Eq.(\ref{eq:ln-det-eq-tr-ln}) implies

\beq
\delta \ln ( \det A) = \tr (A^{-1} \delta A) = \sum_{i, j} (A^{-1})_{ij}\delta A_{ji}
\;.
\eeq
Therefore,

\beq
(A^{-1})_{ij} = \pder{}{A_{ji}} \ln \det A
= \frac{1}{\det A} \pder{(\det A)}{A_{ji}}
\;.
\eeq
This is just the usual expansion of $A^{-1}$ in terms of
cofactors. For definiteness, suppose $A$ is a $3\times 3$ matrix
with columns $\vec{a_1}, \vec{a_2}, \vec{a_3}$.
Suppose $u$ and $v$ are also $3\times 1$ column vectors.
Then

\begin{subequations}
\label{eq:det-A-uv}
\begin{eqnarray}
\lefteqn{
\det (A + uv^T)=
\det[ \vec{a}_1 + v^1 \vec{u}, \vec{a}_2 + v^2 \vec{u}, \vec{a}_2 + v^3 \vec{u}] }\\
&=&
\det A
+ \det[ v^1 \vec{u}, \vec{a}_2 , \vec{a}_ 3]
+ \det[ \vec{a}_1,  v^2 \vec{u} , \vec{a}_ 3]
+ \det[ \vec{a}_1, \vec{a}_2 ,v^ 3 \vec{u}]  \\
&=& \det(A) + \sum_{i,j} u^j\; \pder{(\det A)}{A_{ji}} \;v^i \\
&=& \det(A) ( 1 + v^T A^{-1} u)
\;.
\eeqa
In Eq.(\ref{eq:det-A-uv}), we went from line (a) to (b) by using
the fact that determinants are linear functions of each column.
We also used the fact that determinants with a pair
of proportional columns
are zero, so that, for example,

\beq
\det[ v^1 \vec{u}, v^2 \vec{u} , \vec{a}_ 3]=0
\;.
\eeq
Now setting $u = v/\epsilon$ in Eq.(\ref{eq:det-A-uv})
yields

\beqa
\det (B) &=& \det(A) \left( 1 + \frac{v^T A^{-1} v}{\epsilon} \right)\\
&\approx&\det(A) \left(
\frac{v^T A^{-1} v}{\epsilon} \right)
\;.
\eeqa
QED

\begin{claim}\label{th:gau-ints}
For $x, b \in Reals^{N\times 1}$ and $A\in Reals^{N\times N}$,
define a measure $dG(x)$ so that for any real valued function $f(x)$,

\beq
\int dG(x)\; f(x)=
\prodset{
\int_{-\infty}^{+\infty} dx_j
}{ j \in Z_{1,N} }
\exp \left(
\frac{-x^T A x}{2} + b^T x
\right)
f(x)
\;.
\eeq
Suppose $A$ is a real, positive definite, symmetric matrix.
Suppose $u, v\in Reals^{N\times 1}$ and
$\alpha\in Reals$. Define

\beq
\tilde{A}=
\frac{vv^T}{v^T A^{-1} v}
\;\;, \;\;
B^{-1} = A^{-1} - A^{-1}\tilde{A} A^{-1}
\;.
\eeq
Then

\begin{subequations}
\beq
\int dG(x)\; 1 =
\frac{
{(2\pi)}^{\frac{N}{2}}
}{
\sqrt{\det A}
}
\exp
\left(
\frac{ b^T A^{-1} b}{2}
\right)
\;,
\label{eq:gau-int-1}
\eeq

\beq
\int dG(x)\;  \delta(v^T x) =
[\int dG(x)\; 1] \frac{1}
{
\sqrt{2\pi v^T A^{-1}v}
}
\exp
\left(
\frac{ -b^T A^{-1} \tilde{A} A^{-1}b}{2}
\right)
\;,
\label{eq:gau-int-del}
\eeq

\beq
\int dG(x)\; \theta(u^T x -\alpha\geq 0) =
[ \int dG(x)\; 1] \frac{1}{2}
\erfc
\left[
\frac{
\alpha - u^T A^{-1} b
}{
\sqrt{2 u^T A^{-1} u}
}
\right]
\;,
\label{eq:gau-int-theta}
\eeq

\beq
\int dG(x)\; \delta(v^T x)\theta(u^T x -\alpha\geq 0) =
[\int dG(x)\; \delta(v^T x)]
\frac{1}{2}
\erfc
\left[
\frac{
\alpha - u^T B^{-1}b
}{
\sqrt{2 u^T B^{-1} u}
}
\right]
\;.
\label{eq:gau-int-theta-del}
\eeq
\end{subequations}
\end{claim}

\noindent{\bf proof of Eq.(\ref{eq:gau-int-1}) :}

Since $A$ is symmetric, it can be diagonalized.
By diagonalizing $A$, one can convert $\int dG(x) 1$ into
a product of one dimensional Gaussian integrals.

\noindent{\bf proof of Eq.(\ref{eq:gau-int-del}) :}

For $0<\epsilon <<1$,

\beq
\delta(v^T x) \approx
\frac{
1
}{
\sqrt{2 \pi \epsilon }
}
\exp\left(\frac{-(v^T x)^2}{2\epsilon}\right)
\;.
\eeq
Define $B$ by

\beq
B = A + \frac{v^T v}{\epsilon}
\;.
\eeq
Then

\begin{subequations}
\label{eq:gau-ints}
\begin{eqnarray}
\int dG(x)\; \delta(v^Tx) &= &
\frac{
1
}{
\sqrt{2 \pi \epsilon }
}
\prodset{
\int_{-\infty}^{+\infty} dx_j
}{ j \in Z_{1,N} }
\exp \left(
\frac{-x^T B x}{2} + b^T x
\right)\\
&=&
\frac{
1
}{
\sqrt{2 \pi \epsilon }
}
\frac{
{(2\pi)}^{\frac{N}{2}}
}{
\sqrt{\det B}
}
\exp
\left(
\frac{ b^T B^{-1} b}{2}
\right)
\;.
\eeqa
Now use the values for $B^{-1}$ and $\det B$
calculated in Lemma \ref{lem:inv-det-B}.

\noindent{\bf proof of Eq.(\ref{eq:gau-int-theta}) :}

\begin{subequations}
\label{eq:int-dg-theta}
\begin{eqnarray}
\lefteqn{\int dG(x)\; \theta( u^T x - \alpha \geq 0)
=
\int dG(x) \frac{1}{2\pi i}
\int_{\Lambda - i \infty}^{\Lambda + i \infty}
\frac{d\lam}{\lam}\;
e^{\lam(u^T x - \alpha)}
}\\
&=&
\frac{1}{2 \pi i}
\int_{\Lambda - i \infty}^{\Lambda + i \infty}
\frac{d\lam}{\lam}\;
\int dx^N \exp
\left(
\frac{ - x^T A x }{2}
+ (b + \lam u)^T x - \lam \alpha
\right)\\
&=&
\frac{1}{2 \pi i}
\int_{\Lambda - i \infty}^{\Lambda + i \infty}
\frac{d\lam}{\lam}\;
\frac{
{(2\pi)}^{\frac{N}{2}}
}{
\sqrt{\det A}
}
\exp\left(
\frac{(b + \lam u)^T A^{-1} (b + \lam u)}{2} - \lam \alpha
\right)\\
&=&
\frac{1}{2 \pi i}
\frac{
{(2\pi)}^{\frac{N}{2}}
}{
\sqrt{\det A}
}
\exp\left(
\frac{b^T A^{-1} b}{2}\right)\nonumber\\
&&
\int_{\Lambda - i \infty}^{\Lambda + i \infty}
\frac{d\lam}{\lam}\;
\exp\left(
\lam^2 (\frac{u^T A^{-1} u}{2})
+\lam ( u^T A^{-1} b -\alpha ) \right)\\
&=&
[ \int dG(x)\; 1] \frac{1}{2}
\erfc
\left[
\frac{
\alpha - u^T A^{-1} b
}{
\sqrt{2 u^T A^{-1} u}
}
\right]
\;.
\eeqa
In Eq.(\ref{eq:int-dg-theta}), line (a),
we used the integral representation of the theta function
given by Eq.(\ref{eq:theta-int-rep}).
In Eq.(\ref{eq:int-dg-theta}), we went from  line (b) to (c)
by applying Eq.(\ref{eq:gau-int-1}).
We went  from line (d) to (e) by applying Eq.(\ref{eq:erfc-contour-int-a-b}).

\noindent{\bf proof of Eq.(\ref{eq:gau-int-theta-del}) :}

This proof is similar to that  of
Eqs.(\ref{eq:gau-ints}) (a), (b) and (c) so it is left to the reader.
QED

\section{Appendix: An Integral Over All \\
Joint Probability Distributions\\
 with a Fixed Marginal}\label{app:int-fixed-margi}

In this appendix, we will show how to convert (1) to (2) where
(1) is an integral over all joint probability distributions $P_{\rvx, \rvy}$
with the same marginal $P_\rvy$, and (2) is
an integral over all conditional probability distributions
$P_{\rvx|\rvy}$.

\begin{claim}
\begin{eqnarray}\label{eq:dpxy-del-y}
\lefteqn{\int \calD P_{\rvx, \rvy}\;
\prodset{\delta(P(y) - Q(y))}{y}
\theta(P_{\rvx, \rvy}\geq 0)
f(P_{\rvx, \rvy}) =
}\nonumber \\
&&
\prodset{[Q(y)]^{N_\rvx-1}}{y}
\int \prodset{dP(x|y)}
{x, y}\nonumber \\
&&
\prodset{\delta(\sum_x P(x|y) -1)}
{y}
\theta(P_{\rvx|\rvy}\geq 0)
f(P_{\rvx, \rvy})
\;.
\end{eqnarray}
\end{claim}
{\bf proof:}

Let RHS (ditto, LHS)
stand for the right (ditto, left) hand side of Eq.(\ref{eq:dpxy-del-y}).
Suppose $0\in S_\rvx$. Then

\beqa
LHS
&=& \int
\prodset{  dP(x, y)}{(x, y), x\neq 0}\nonumber \\
&&
\prodset{
\theta[
0 \leq \sum_{x:\; x\neq 0} P(x,y) \leq Q(y)
]
}{y}
\theta(P_{\rvx, \rvy}> 0) f(P)\\
&=&
\prodset{[Q(y)]^{N_\rvx-1}}{y}
\int \prodset{dP(x|y)}
{(x, y), x\neq 0}\nonumber\\
&&
\prodset{
\theta(0 \leq \sum_{x:\; x\neq 0} P(x|y) \leq 1)
}{y}
\theta(P_{\rvx| \rvy}\geq 0)
f(P)\\
&=& RHS
\;.
\eeqa
QED

\section{Appendix: Perturbation Expansion of $t$}\label{app:t-exp}

In Eq.(\ref{eq:leading-t}), we gave $t$ to lowest order in $\Delta P$.
In this appendix, we show how to calculate $t$ exactly, as a Taylor
series in powers of $\Delta P$.

The point $\pwave^*$
that dominates the integral Eq.(\ref{eq:f-integral-2})
is an extremum of the Lagrangian Eq.(\ref{eq:lagr-noisy-exact}).
In Section \ref{sec:new-noisy-perr}, we approximated the
Lagrangian Eq.(\ref{eq:lagr-noisy-exact}) by
its quadratic approximation  Eq.(\ref{eq:lagr-noisy-quad}).
This gave us the dominant point $\pwave^*$ only to lowest
order in $\Delta P$. This time we will use the exact Lagrangian
and get the exact dominant point. Let us re-state the
exact Lagrangian:

\beq
\calL = D(\pwave_{\rvx, \rvy}// Q_{\rvx, \rvy})
-\lam \left ( \sum_{x, y} (P - \pwave)(x, y) L_{xy} \right)
+\sum_y \mu_y ( P - \pwave)(y)
\;.
\label{eq:lagr-noisy-exact2}
\eeq
Minimizing this Lagrangian with respect to
$\pwave$, $\lam$ and $\mu_y$ gives

\beq
\pwave^*(x, y) =
\frac{Q(x|y) \exp(-\lam L_{xy})
}{
Z_y(\lam)
}P(y)
\;,
\label{eq:pwave-lam}
\eeq
where

\beq
Z_y(\lam) = \sum_x Q(x|y)  \exp(-\lam L_{xy})
\;.
\label{eq:zy-def}
\eeq
The parameter $\lam$ in Eq.(\ref{eq:pwave-lam})
is specified implicitly by the equation:

\beqa
\sum_{x,y} P(x,y) L_{xy}
&=&
\sum_{x,y}
\frac{Q(x|y) \exp(-\lam L_{xy})
}{
Z_y(\lam)
}P(y)L_{xy}\\
&=&
-\sum_y P(y)
\frac{d \ln Z_y (\lam)
}{
d\lam}
\;.
\eeqa
The previous equation can be rewritten as

\beq
0=
\epsilon + F(\lam)
\;,
\label{eq:eps-plus-F}
\eeq
where $\epsilon$ and $F(\lam)$ are defined by

\beq
\epsilon = \sum_{x,y} P(y) \Delta P(x|y) L_{xy}
\;,
\eeq
and

\beq
F(\lam)
= \sum_y P(y)
\left[
\frac{ d\ln Z_y (\lam)
}{
d\lam}
-
\left(
\frac{d \ln Z_y (\lam)
}{
d\lam}
\right)_{\lam = 0}
\right]
\;.
\label{eq:F-eq-dlnzy}
\eeq

Next we will solve Eq.(\ref{eq:eps-plus-F})
for $\lam$ by expressing $\lam$
as a Taylor series in powers of $\epsilon$. We begin by
expressing the RHS of Eq.(\ref{eq:zy-def})
as a Taylor series in powers of  $\lam$:

\beq
Z_y(\lam) =\sum_{k=0}^{\infty}
\frac{ A_k(y) (-\lam)^k}{k!}
\;,
\eeq
where

\beq
A_k(y) = \sum_x Q(x|y)(L_{xy})^k
\;
\eeq
for $k=0, 1, 2, \cdots$. It follows that

\beq
\ln Z_y(\lam) =
a_1\lam + a_2 \frac{\lam^2}{2} + a_3 \frac{\lam^3}{3} + \ldots
\;,
\label{eq:lnzy-expan}
\eeq
where

\begin{subequations}
\beq
a_1 = -A_1
\;,
\eeq

\beq
a_2 = - A_1^2 + A_2
\;,
\eeq

\beq
a_3 = -A_1^3 + \frac{3}{2} A_1 A_2 - \frac{1}{2}A_3
\;,
\eeq

\beq
a_4 = -A_1^4 + 2 A_2 A_1^2 - \frac{2}{3} A_1 A_3
-\frac{1}{2}A_2^2
+\frac{1}{6}A_4
\;.
\eeq
\end{subequations}
Define

\beq
\alpha_k = \sum_y P(y) a_k(y)
\;
\label{eq:alpk-def}
\eeq
for $k=1, 2, 3, \ldots$.
If we express $F(\lam)$ as a Taylor series in powers of $\lam$

\beq
F(\lam)=F_1 \lam
 + F_2 \lam^2 + F_3 \lam^3 + \ldots
\;,
\eeq
then, by virtue of Eqs.(\ref{eq:F-eq-dlnzy}),
(\ref{eq:lnzy-expan}) and (\ref{eq:alpk-def}),

\beq
F_k = \alpha_{k+1}
\;
\eeq
for $k=1, 2, 3, \ldots$.
Eq.(\ref{eq:eps-plus-F}) can be
expressed as a Taylor series in powers of $\lam$:

\beq
0 = \epsilon + F_1 \lam
 + F_2 \lam^2 + F_3 \lam^3 + \ldots
\;.
\label{eq:eps-plus-F-expan}
\eeq
$\lam$ itself can be expressed as a Taylor series in powers of $\epsilon$:

\beq
\lam = \lam_1 \epsilon
 + \lam_2 \epsilon^2
 + \lam_3 \epsilon^3
 + \ldots
\;.
\label{eq:lam-expan}
\eeq
Substituting Eq.(\ref{eq:lam-expan})
into Eq.(\ref{eq:eps-plus-F-expan})
yields an equation for each power of $\epsilon$.
These equations for each power of $\epsilon$
imply:

\begin{subequations}
\beq
\lam_1 =
\frac{-1}{F_1}
\;,
\eeq

\beq
\lam_2 =
\frac{-F_2}{F_1^3}
\;,
\eeq

\beq
\lam_3  =
\frac{ F_3 F_1 - 2F_2^2}
{ F_1^5}
\;,
\eeq

\beq
\lam_4  =
\frac{ -5F_2^3 + 5F_3F_2 F_1 - F_4 F_1^2}
{ F_1^7}
\;.
\eeq
\end{subequations}

Now that we know $\pwave^*_{\rvx, \rvy}$
explicitly
(in terms of Eq.(\ref{eq:pwave-lam}), where $\lam$ is
expressed as a Taylor series in powers of $\epsilon$),
we can find explicitly
$\calL$ given by Eq.(\ref{eq:lagr-noisy-exact2})
evaluated at $\pwave^*_{\rvx,\rvy}$ .

\beqa
\calL^*
 &=&
D(\pwave^*_\rvy//Q_\rvy)
+
\sum_{x,y} \pwave^*(x, y) \ln \left(
\frac{
\pwave^*(x| y)
}{
Q(x|y)
}
\right)
\\
&=&
D(P_\rvy//Q_\rvy)
+
\sum_{x,y} \pwave^*(x, y) \ln \left(
\frac{
\exp(-\lam L_{xy})
}{
Z_y(\lam)
}
\right)
\\
&=&
D(P_\rvy//Q_\rvy)
- \lam \sum_{x,y} P(x,y) L_{xy}
- \sum_y P(y) \ln Z_y(\lam)
\;.
\eeqa
Expanding the $\ln Z_y(\lam)$ in the previous equations
in powers of $\lam$ yields

\beqa
\calL^* &=&
D(P_\rvy//Q_\rvy)
- \lam \sum_{x,y} P(x,y) L_{xy}\nonumber \\
&& -(
\lam \alpha_1
+ \lam^2 \frac{\alpha_2}{2}
+ \lam^3 \frac{\alpha_3}{3}
+ \cdots
)\\
&=&
D(P_\rvy//Q_\rvy)
-(
\lam \epsilon
+ \lam^2 \frac{\alpha_2}{2}
+ \lam^3 \frac{\alpha_3}{3}
+ \cdots
)
\;.
\eeqa
Expanding $\lam$ in the previous equation in powers
of $\epsilon$ yields

\beq
\calL^*=
D(P_\rvy//Q_\rvy)
+ t
\;,
\eeq
where

\beq
t =
 t_1\epsilon
+ t_2\epsilon^2
+ t_3\epsilon^3
+ \cdots
\;,
\label{eq:t-epsilon-expan}
\eeq
and

\begin{subequations}
\label{eq:t-exp-coef}
\beq
t_1 = 0
\;,
\eeq

\beq
t_2 = \frac{1}{2 \alpha_2}
\;,
\eeq

\beq
t_3 =
\frac{ \alpha_3}{ 3 \alpha_2^3}
\;,
\eeq

\beq
t_4 =
\frac{
2 \alpha_3^2 - \alpha_4\alpha_2
}{
4 \alpha_2^5
}
\;.
\eeq
\end{subequations}

Now that we know $t$ to all orders in $\epsilon$,
we can also find $\gamma_1$ to all orders in $\epsilon$.
Recall from Section \ref{sec:new-noisy-rg-eqs}
that for any real valued function $f(s)$ of $s\geq 0$,

\beq
Df = \lim_{s\rarrow 0} \left(\frac{-1}{\gamma_0}\right)\pder{f}{s}
\;,
\eeq
so  that $DP^\sfun = \Delta P$. When $f$ is the $k$th power of
$\epsilon$,

\beqa
D\epsilon^k &=&
 k \epsilon^{k-1} \sum_{x,y} [\Delta P(x,y) - Q(x|y)\Delta P(y)]L_{xy}\\
&=& k \epsilon^k
\;.
\eeqa
From Eq.(\ref{eq:t-epsilon-expan})
one gets

\beq
Dt =
\left\{
\begin{array}{l}
t_1 \epsilon + t_2 2 \epsilon^2 + t_3 3 \epsilon^3 +\ldots\\
+ \epsilon Dt_1 + \epsilon^2 Dt_2 + \epsilon^3 Dt_3 + \ldots
\end{array}
\right.
\;.
\eeq
One can use Eqs.(\ref{eq:t-exp-coef})
to calculate $Dt_k$ in terms of $\{ \alpha_k \}_{\forall k}$
and $\{ D\alpha_k \}_{\forall k}$.
For example, $Dt_2 = \frac{-1}{2 \alpha^2_2} D\alpha_2$.
By Eq.(\ref{eq:alpk-def}),

\beq
D\alpha_k = \sum_y \Delta P(y) a_k(y)
\;,
\eeq
for $k = 1, 2, 3 , \ldots$.
Once we know $t$ and $Dt$ to all orders in $\epsilon$,
we can use Eq.(\ref{eq:gamma-one-from-t-Dt})
 to find $\gamma_1$ to all orders in $\epsilon$.


\begin{thebibliography}{99}
\bibitem{Gold}Nigel Goldenfeld, {\it Lectures on Phase
Transitions and the Renormalization Group} (1992, Perseus Books).

\bibitem{cover-thomas}T.M. Cover, J.A. Thomas,
{\it Elements of Information Theory} (1991, John Wiley).

\bibitem{blahut}R.E. Blahut,
{\it Principles and Practice of Information Theory} (1987, Addison-Wesley)


\bibitem{asymp}G.F. Carrier, M. Krook, C.E. Pearson,
{\it Functions of a Complex Variable} (1966, MacGraw-Hill);
N. Bleistein, R. A. Handelsman, {\it Asymptotic
Expansions of Integrals} (1986, Dover).


\bibitem{Fletcher}R. Fletcher,
{\it Practical Methods of Optimization}
(2000, John Wiley).

\bibitem{first-Del-R}
An alternative method of getting a good trial value for
$\Delta R^\sfinfun$ is as follows.
Note that $\gamma_0(P, Q)$ and $\gamma_1(P, Q)$
both tend to $\frac{1}{2}$ as $P\rarrow Q$.
Thus, a good trial value for $\Delta R^\sfinfun$ is
$e^{\frac{-\sfin}{2}} \Delta R$.
Plug this value of $\Delta R^\sfinfun$ into
$\Delta P(X) = B(X)\Delta R^\sfinfun$
and check that it gives $P^\sfinfun(X) \in [0,1]$ for all $X$.
If not, then continue halving the trial value of
$\Delta R^\sfinfun$ until
$P^\sfinfun(X)\in[0,1]$
for all $X$. This  occurs eventually, assuming
$Q(X) \neq 0$ for all $X$.


\bibitem{AbraSteg}
M. Abramowitz, I.A. Stegun,
{\it Handbook of Mathematical Functions} (1974, Dover).


\end{thebibliography}
\end{document}